\documentclass[aps,prd,10pt,nofootinbib]{revtex4}
\usepackage{graphicx}
\usepackage{amssymb}
\usepackage{hyperref}
\hypersetup{colorlinks=true,linkcolor=blue,citecolor=red,anchorcolor=black}
\begin{document}
\title{Goldstone bosons and the Englert-Brout-Higgs mechanism in non-Hermitian theories}
\author{Philip D. Mannheim}
\affiliation{Department of Physics, University of Connecticut, Storrs, CT 06269, USA\\
email: philip.mannheim@uconn.edu}
\date{January 17, 2019}

\begin{abstract}
In recent work Alexandre, Ellis, Millington and Seynaeve have extended the Goldstone theorem to non-Hermitian Hamiltonians that possess a discrete antilinear symmetry such as $PT$ and possess a continuous global symmetry. They restricted their discussion to those realizations of antilinear symmetry in which all the energy eigenvalues of the Hamiltonian are real. Here we extend the discussion to the two other realizations possible with antilinear symmetry, namely energies in complex conjugate pairs or Jordan-block Hamiltonians that are not diagonalizable at all. In particular, we show that under certain circumstances it is possible for the Goldstone boson mode itself to be one of the  zero-norm states that  are characteristic of Jordan-block Hamiltonians. While we discuss the same model as Alexandre, Ellis, Millington and Seynaeve our treatment is quite different, though their main conclusion  that one can have Goldstone bosons in the non-Hermitian case remains intact. We extend our analysis to a continuous local symmetry and find that the gauge boson acquires a non-zero mass by the Englert-Brout-Higgs mechanism in all realizations of the antilinear symmetry, except the one where the Goldstone boson itself has zero norm, in which case, and despite the fact that the continuous local symmetry has been spontaneously broken,  the gauge boson remains massless.
\end{abstract}
\maketitle

\section{Introduction}
\label{S1}

Following work by Bender and collaborators \cite{Bender1998,Bender1999,Bender2007,Special2012,Theme2013} it has become apparent that quantum mechanics is much richer than conventional Hermitian quantum mechanics. However, if one wishes to maintain probability conservation, one needs to be able to define an inner product that is time independent. The reason that one has any freedom at all in doing this is because the Schr\"odinger equation $i\partial_t|\psi\rangle= H|\psi\rangle$ only involves the ket state and leaves the bra state unspecified. While the appropriate bra state is the Hermitian conjugate of the ket when the Hamiltonian is Hermitian, for non-Hermitian Hamiltonians a more general bra state is needed. However, one cannot define an inner product that is time independent for any non-Hermitian Hamiltonian. Rather, it has been found (\cite{Mannheim2013,Mannheim2018} and references therein) that the most general Hamiltonian for which one can construct a time-independent inner product is one that has an antilinear symmetry, and in such a case the required bra state is the conjugate of the ket state with respect to that particular antilinear symmetry. 

When a Hamiltonian has an antilinear symmetry its energy eigenspectrum can be realized in three possible ways, all eigenvalues real and eigenspectrum complete, some or all of the eigenvalues in complex conjugate pairs with the eigenspectrum still being complete, or eigenspectrum incomplete  and Hamiltonian of non-diagonalizable, and thus necessarily of non-Hermitian,  Jordan-block form. Of these three possible realizations only the first can also be achieved with a Hermitian Hamiltonian, and while Hermiticity implies the reality of energy eigenvalues, there is no theorem that would require a non-Hermitian Hamiltonian to have complex eigenvalues, with Hermiticity only being sufficient for the reality of eigenvalues but not necessary.\footnote{For any non-diagonalizable two-dimensional Jordan-block and thus necessarily non-Hermitian Hamiltonian for instance, since the eigenspectrum is incomplete the Hamiltonian has just one eigenvector even though there are two eigenvalue solutions to $|H-\lambda I|=0$. These two eigenvalue solutions must then be equal to each other since they have to share just the one eigenvector. If in addition the Hamiltonian has an antilinear symmetry, by being equal to each other the two eigenvalues could then not be in a complex conjugate pair. In consequence, the two eigenvalue solutions to $|H-\lambda I|=0$  must be real -- to thus show directly that one can have real eigenvalues if a Hamiltonian is not Hermitian.} The necessary condition for the reality of energy eigenvalues is that the Hamiltonian have an antilinear symmetry \cite{Mostafazadeh2002,Solombrino2002,Bender2010,Mannheim2018}, while the necessary and sufficient condition is that in addition all energy eigenstates are eigenstates of the antilinear operator \cite{Bender2007}.

Interest in non-Hermitian Hamiltonians with an antilinear symmetry was first triggered by the work of Bender and collaborators \cite{Bender1998,Bender1999}  who found that the eigenvalues of the Hamiltonian $H=p^2+ix^3$ are all real. This surprising reality was traced to the fact that the Hamiltonian possesses an antilinear $PT$ symmetry ($P$ is parity and $T$ is time reversal), under which $PpP=-p$, $PxP=-x$, $TpT=-p$, $TxT=x$, $TiT=-i$. In general for any Hamiltonian $H$ with an antilinear symmetry $A$ (i.e. with $AH=HA$), when acting  on $H|\psi\rangle=E|\psi\rangle$ one has $AH|\psi\rangle =AHA^{-1}A|\psi\rangle=HA|\psi\rangle=E^*A|\psi\rangle$. Thus for every eigenstate $|\psi\rangle$ of $H$ with energy $E$ there is another eigenstate  $A|\psi\rangle$ of $H$  with energy $E^*$. Thus as originally noted by Wigner in his study of time reversal invariance,  energies can thus be real or appear in complex conjugate pairs with complex conjugate eigenfunctions. It is often the case that one can move between these two realizations by a change in the parameters in $H$. There will thus be a transition point (known as an exceptional point) at which the switch over occurs. However, at this transition point the two complex conjugate wave functions ($|\psi\rangle$ and $A|\psi\rangle$) have to collapse into a single common wave function as there are no complex conjugate pairs on the real energy side. Since this collapse to a single common wave function reduces the number of energy eigenfunctions, at the transition point the eigenspectrum of the Hamiltonian becomes incomplete, with the Hamiltonian then being of non-diagonalizable Jordan-block form, the thus third possible realization of antilinear symmetry.

While the above analysis would in principle apply to any antilinear symmetry, because of its $H=p^2+ix^3$ progenitor, the antilinear symmetry program is conventionally referred to as the $PT$-symmetry program. However, $PT$ symmetry can actually be selected out for a different reason, namely it has a connection  to spacetime. Specifically, it was noted in \cite{Streater1964} and emphasized in \cite{Bender2007} that for the spacetime coordinates the linear part of a  $PT$ transformation is the same as a particular complex Lorentz transformation, while in \cite{Mannheim2016,Mannheim2018} it was noted that for spinors the linear part of a $CPT$ transformation is the same as that very same particular complex Lorentz transformation, where $C$ denotes charge conjugation.\footnote{The complex Lorentz transformation $\Lambda^{0}_{\phantom{0}3}(i\pi)\Lambda^{0}_{\phantom{0}2}(i\pi)\Lambda^{0}_{\phantom{0}1}(i\pi)$ implements $x_{\mu}\rightarrow -x_{\mu}$ on coordinates and $\psi_1(x) \rightarrow\gamma^5 \psi_1(-x)$ on a Majorana spinor, just as the linear part of a $CPT$ transformation does.}   Then in \cite{Mannheim2016,Mannheim2018} it was shown that if one imposes only two requirements, namely the time independence of  inner products and invariance under the complex Lorentz group, it follows that the Hamiltonian  must be $CPT$ invariant, with $CPT$ symmetry itself being antilinear. Since this analysis involves no Hermiticity requirement, the $CPT$ theorem is thus extended to the non-Hermitian case (and thus through the  complex energy realization of antilinear symmetry to decay processes that are forbidden by Hermiticity). Since charge conjugation plays no role in non-relativistic physics where one is below the threshold for particle production, $CPT$ then defaults to $PT$, to thus put the $PT$-symmetry program on a quite secure theoretical foundation.

As with the $CPT$ theorem, one can ask what happens to other familiar results of quantum field theory when one relaxes the Hermiticity requirement. This then was the brief of Alexandre, Ellis, Millington and Seynaeve \cite{Alexandre2018}, who found that the Goldstone theorem can also be decoupled from Hermiticity, and can hold in the non-Hermitian but antilinearly symmetric case.\footnote{Since historically the $CPT$ theorem was found during the effort to establish the spin and statistics theorem, it would be of interest to see how the spin and statistics theorem itself might fare in the non-Hermitian but $CPT$ symmetric case.}  Alexandre, Ellis, Millington and Seynaeve restricted their discussion to those realizations of antilinear symmetry in which all the energy eigenvalues of the Hamiltonian are real. Here we extend the discussion to the two other possible $PT$-symmetry program realizations, namely energies in complex conjugate pairs or Jordan-block Hamiltonians that are not diagonalizable at all. In particular, we show that it is possible for the Goldstone boson mode itself to be one of the  zero-norm states that  are characteristic of Jordan-block Hamiltonians. While we discuss the same model as Alexandre, Ellis, Millington and Seynaeve our treatment is quite different, though their main conclusion  that one can have Goldstone bosons in the non-Hermitian case remains intact. In particular, in their paper Alexandre, Ellis, Millington and Seynaeve presented a variational procedure for the action in which the surface term played an explicit role, to thus suggest that one has to use such a procedure in order to establish the Goldstone theorem in the non-Hermitian case. However, we show that one does not need to do this, as we are able to obtain a Goldstone boson using a completely standard variational procedure. Moreover, since we do use a standard variational procedure we can readily extend our analysis to a continuous local symmetry by introducing a gauge boson. We show that the gauge boson acquires a non-zero mass by the Englert-Brout-Higgs mechanism in all realizations of the antilinear symmetry, except the one where the Goldstone boson itself has zero norm, in which case, and despite the spontaneous breakdown of the continuous local symmetry, the gauge boson remains massless.

The present paper is organized as follows. In Sec. \ref{S2} we present  the complex scalar field model discussed in \cite{Alexandre2018}, and using a standard variational procedure for the action find its spontaneously broken tree approximation minimum and determine  the eigenvalues of the associated mass matrix. In Sec. \ref{S3} we determine the associated left- and right-eigenvectors and construct the left-right $V$ operator norm that plays a central role in antilinear theories. In Sec. \ref{S4} we compare our treatment with that of the authors of \cite{Alexandre2018}, who used a non-standard variational procedure. This leads us to a Hamiltonian that looks Hermitian but is not, and in Sec. \ref{S5} we discuss how this is possible. In this section we also discuss the connection between antilinear symmetry and Hermiticity within the context of the $CPT$ theorem as developed in  \cite{Mannheim2018}. In Sec. \ref{S6} we extend the discussion to the Englert-Brout-Higgs mechanism, and in Sec. \ref{S7} we provide a summary of our results. Finally, in an appendix we construct the left-right quantum theory matrix elements that would produce the c-number tree approximation classical field and the effective potential that is minimized in Sec. \ref{S2}, and discuss how Ward identities are realized in the non-Hermitian case.

\section{Spontaneously Broken non-Hermitian Theory with a Continuous Global Symmetry} 
\label{S2}

The model introduced in \cite{Alexandre2018} consists of two complex (i.e. charged) scalar fields $\phi_1(x)$ and $\phi_2(x)$ with action 
\begin{eqnarray}
I(\phi_1,\phi_2,\phi^*_1,\phi^*_2)=\int d^4x\left[\partial_{\mu}\phi^*_1\partial^{\mu}\phi_1+\partial_{\mu}\phi^*_2\partial^{\mu}\phi_2
+m_1^2\phi_1^*\phi_1-m_2^2\phi^*_2\phi_2-\mu^2(\phi^*_1\phi_2-\phi^*_2\phi_1)-\frac{g}{4}(\phi^*_1\phi_1)^2\right],
\label{GPT1}
\end{eqnarray}
where the star symbol denotes complex conjugation, and thus Hermitian conjugation since neither of the the two scalar fields possesses any internal symmetry index. Since the action is not invariant under complex conjugation, it is not Hermitian. It is however invariant under the following $CPT$ transformation
\begin{eqnarray}
\phi_1(x_{\mu})\rightarrow \phi^*_1(-x_{\mu}),\quad \phi_2(x_{\mu})\rightarrow -\phi^*_2(-x_{\mu}),
\quad \phi^*_1(x_{\mu})\rightarrow \phi_1(-x_{\mu}),\quad \phi^*_2(x_{\mu})\rightarrow -\phi_2(-x_{\mu}),
\label{GPT2}
\end{eqnarray}
and thus has an antilinear symmetry.\footnote{The study of \cite{Mannheim2018,Mannheim2016} shows that for relativistic actions such as that given in (\ref{GPT1})  $CPT$ must be an invariance, a point we elaborate on further below. In their paper the authors of \cite{Alexandre2018} took $T$ to conjugate fields. While $T$ does conjugate wave functions in quantum mechanics, conventionally in quantum field theory $T$ does not conjugate q-number fields (it only conjugates c-numbers). Rather, it is charge conjugation $C$ that conjugates fields. Thus what the authors of \cite{Alexandre2018} refer to as $PT$ is actually $CPT$, just as required by the analysis of \cite{Mannheim2018,Mannheim2016}. However none of the conclusions of \cite{Alexandre2018} are affected by this.} Since one can construct the energy-momentum tensor $T_{\mu\nu}$ by the variation $T^{\mu\nu}=2(-g)^{-1/2}\delta I(\phi_1,\phi_2,\phi_1^*,\phi_2^*)/\delta g_{\mu\nu}$ with respect to the metric $g_{\mu\nu}$ of the covariantized form of the action (momentarily replace ordinary derivatives by covariant ones and replace the measure by $\int d^4x (-g)^{1/2}$), it follows from general coordinate invariance that  a so-constructed energy-momentum tensor is automatically covariantly conserved in solutions to the equations of motion that follow from stationarity of the same action. Then, since one can set $H=\int d^3x T_{00}$, it follows that the associated Hamiltonian is time independent. Moreover, since the metric is $CPT$ even, then since the action is $CPT$ invariant it follows that  the Hamiltonian is $CPT$ invariant too. The Hamiltonian associated with (\ref{GPT1}) thus has an antilinear $CPT$ symmetry.\footnote{Just as is familiar from Hermitian quantum field theory,  one can also construct the same metric-derived energy-momentum tensor from the translation invariance of the action and the equations of motion of the fields, since nothing in that construction actually requires Hermiticity. The advantage of using the metric approach, which also is not sensitive to Hermiticity, is that it ensures that the Hamiltonian that is obtained has the same transformation properties under $CPT$ symmetry as the starting action.}

In regard to (\ref{GPT2}), we note here that for $\phi_2$ the transformation is not the conventional $CPT$ transformation of scalar fields that is used in quantum field theory (one in which all scalar field $CPT$ phases are positive \cite{Weinberg1995}) but a similarity transformation of it. We will need to return to this point below, but for the moment we just use (\ref{GPT2}) as is. 

As written, the action given in (\ref{GPT1}) is invariant under the electric charge transformation 
\begin{eqnarray}
\phi_1\rightarrow e^{i\alpha}\phi_1,\quad \phi^*_1\rightarrow e^{-i\alpha}\phi^*_1,\quad \phi_2\rightarrow e^{i\alpha}\phi_2,\quad \phi^*_2\rightarrow e^{-i\alpha}\phi_2,
\label{GPT3}
\end{eqnarray}
to thus possess a standard Noether current
\begin{eqnarray} 
j_{\mu}=i(\phi^*_1\partial_{\mu} \phi_1-\phi_1\partial_{\mu} \phi^*_1)+i(\phi^*_2\partial_{\mu} \phi_2-\phi_2\partial_{\mu} \phi^*_2)
\label{GPT4}
\end{eqnarray}
that is conserved in solutions to the equations of motion {(\ref{GPT36}) and (\ref{GPT37})} associated with (\ref{GPT1}). We note here that the authors of \cite{Alexandre2018} used a non-standard Euler-Lagrange variational procedure (one which involves a non-trivial surface term) to obtain a non-standard set of equations of motion and a non-standard current (one not a Noether current invariance of the action), one that is nonetheless conserved in solutions to this non-standard set of equations of motion, and we discuss this issue in Sec. \ref{S4}. However, we shall use a standard variational procedure and a standard Noether current approach. With the potential of the field $\phi_1$ being of the form of a double-well potential, in its non-trivial minimum the scalar field $\phi_1$ would acquire a non-trivial vacuum expectation value. This would then break the electric charge symmetry spontaneously, and one would thus wonder whether there might still be a massless Goldstone boson despite the lack of Hermiticity. As shown by the authors of \cite{Alexandre2018} for the current they use  and by us here for the above $j_{\mu}$, in both the cases a Goldstone boson is indeed present.

To study the dynamics associated with the action given in (\ref{GPT1}) we have found it convenient to work in the component basis
\begin{eqnarray} 
\phi_1=\frac{1}{\sqrt{2}}(\chi_1+i\chi_2),\quad \phi^*_1=\frac{1}{\sqrt{2}}(\chi_1-i\chi_2),\quad \phi_2=\frac{1}{\sqrt{2}}(\psi_1+i\psi_2),\quad \phi^*_2=\frac{1}{\sqrt{2}}(\psi_1-i\psi_2),
\label{GPT5}
\end{eqnarray}
where all four $\chi_1$, $\chi_2$, $\psi_1$,  and $\psi_2$ are Hermitian.\footnote{As is standard, under time reversal $\chi_1$ has even $T$ parity while $\chi_2$ has odd $T$ parity, so that under $T$ $\chi_1+i\chi_2$ has even parity. Under charge conjugation, $\chi_1$ has even $C$ parity while $\chi_2$ has odd $C$ parity. Thus under $CPT$ the $P$ even $\chi_1+i\chi_2$ transforms into $\chi_1-i\chi_2$. Because of the transformations in the $\phi_2$ sector that are given in (\ref{GPT2})  $\psi_1$ has to have odd $T$ parity while $\psi_2$ has to have even $T$ parity. (However, their $C$ parities are standard, with $\psi_1$ having even $C$ parity while $\psi_2$ has odd $C$ parity.) We discuss this pattern of $T$ parity assignments further below, where we will make a commutation relation preserving similarity transformation that will effect $\psi_1\rightarrow -i\psi_1$,  $\psi_2\rightarrow -i\psi_2$, to thus change the signs of their  $T$ and $CPT$ parities.} In the $\chi_1$, $\chi_2$, $\psi_1$,  and $\psi_2$ basis the action takes the form:
\begin{eqnarray} 
I(\chi_1,\chi_2,\psi_1,\psi_2)&=&\int d^4x \bigg{[}\frac{1}{2}\partial_{\mu}\chi_1\partial^{\mu}\chi_1+
\frac{1}{2}\partial_{\mu}\chi_2\partial^{\mu}\chi_2+
\frac{1}{2}\partial_{\mu}\psi_1\partial^{\mu}\psi_1+
\frac{1}{2}\partial_{\mu}\psi_2\partial^{\mu}\psi_2+
\frac{1}{2}m_1^2(\chi_1^2+\chi_2^2)
\nonumber\\
&-&
\frac{1}{2}m_2^2(\psi_1^2+\psi_2^2)-
i\mu^2(\chi_1\psi_2-\chi_2\psi_1)
-\frac{g}{16}(\chi_1^2+\chi_2^2)^2
\bigg{]},
\label{GPT6}
\end{eqnarray}
and with the appearance of the factor $i$ in the $\mu^2$-dependent term, the action now has the characteristic form of the non-Hermitian but $PT$ symmetric $p^2+ix^3$ theory.\footnote{The  $PT$-symmetric $p^2+ix^3$ theory is actually $CPT$ symmetric since $p$ and $x$ are $C$ even and charge conjugation plays no role in non-relativistic systems.} For this action the Euler-Lagrange equations of motion take the form
\begin{eqnarray}
-\Box \chi_1&=&-m_1^2\chi_1+i\mu^2\psi_2+\frac{g}{4}(\chi_1^3+\chi_1\chi_2^2),
\nonumber\\
-\Box \chi_2&=&-m_1^2\chi_2-i\mu^2\psi_1+\frac{g}{4}(\chi_2^3+\chi_2\chi_1^2),
\nonumber\\
-\Box \psi_1&=&m_2^2\psi_1-i\mu^2\chi_2,
\nonumber\\
-\Box \psi_2&=&m_2^2\psi_2+i\mu^2\chi_1,
\label{GPT7}
\end{eqnarray}
and admit of a tree approximation minimum in which the scalar field expectation values obey 
\begin{eqnarray}
&&m^2_2\bar{\psi}_1-i\mu^2\bar{\chi}_2=0,\quad m_2^2\bar{\psi}_2+i\mu^2\bar{\chi}_1=0,
\nonumber\\
&&m_1^2\bar{\chi}_1-\frac{\mu^4}{m_2^2}\bar{\chi}_1-\frac{g}{4}\bar{\chi}_1^3-\frac{g}{4}\bar{\chi}_1\bar{\chi}_2^2=0,
\nonumber\\
&&m_1^2\bar{\chi}_2-\frac{\mu^4}{m_2^2}\bar{\chi}_2-\frac{g}{4}\bar{\chi}_2^3-\frac{g}{4}\bar{\chi}_2\bar{\chi}_1^2=0.
\label{GPT8}
\end{eqnarray}
Choosing the minimum in which $(g/4)\bar{\chi}_1^2=m_1^2-\mu^4/m_2^2$, $\bar{\psi}_2=-i\mu^2\bar{\chi}_1/m_2^2$, $\hat{\chi}_2=0$, $\hat{\psi}_1=0$, and then expanding around this minimum according to $\chi_1=\bar{\chi}_1+\hat{\chi}_1$, $\chi_2=\hat{\chi}_2$, $\psi_1=\hat{\psi}_1$, $\psi_2=\bar{\psi}_2+\hat{\psi}_2$ yields a first-order term in the equations of motion of the form:
\begin{eqnarray}
\pmatrix{-\Box \hat{\chi}_1 \cr -\Box \hat{\psi}_2 \cr -\Box \hat{\chi}_2 \cr -\Box \hat{\psi}_1 }=
\pmatrix{ 2m_1^2-3\mu^4/m_2^2 &  i\mu^2 & 0&0 \cr 
i\mu^2& m_2^2&0&0\cr
0&0&-\mu^4/m_2^2&-i\mu^2\cr
0&0&-i\mu^2&m_2^2}
\pmatrix{ \hat{\chi}_1 \cr  \hat{\psi}_2 \cr  \hat{\chi}_2 \cr  \hat{\psi}_1 }= M\pmatrix{ \hat{\chi}_1 \cr  \hat{\psi}_2 \cr  \hat{\chi}_2 \cr  \hat{\psi}_1 }.
\label{GPT9}
\end{eqnarray}
As we see, with our choice of basis, we have already block-diagonalized the mass matrix $M$. We can readily determine the mass eigenvalues, and obtain
\begin{eqnarray}
|M-\lambda I|=\lambda (\lambda +\mu^4/m_2^2-m_2^2)\left[\lambda^2 -\lambda (2m_1^2+m_2^2-3\mu^4/m_2^2)+2m_1^2m_2^2-2\mu^4\right].
\label{GPT10}
\end{eqnarray}
The mass eigenvalue solutions to $|M-\lambda I|=0$  are thus 
\begin{eqnarray}
\lambda_0=0,\quad \lambda_1&=&\frac{m_2^4-\mu^4}{m_2^2},
\nonumber\\
\lambda_{\pm}&=&\frac{2m_1^2m_2^2+m_2^4- 3\mu^4}{2m_2^2} \pm
\frac{1}{2m_2^2}\left[(2m_1^2m_2^2+m_2^4-3\mu^4)^2+8\mu^4m_2^4-8m_1^2m_2^6\right]^{1/2}.
\nonumber\\
&=&\frac{2m_1^2m_2^2+m_2^4-3\mu^4}{2m_2^2} \pm
\frac{1}{2m_2^2}\left[(2m_1^2m_2^2-m_2^4-3\mu^4)^2-4\mu^4m_2^4\right]^{1/2}.
\label{GPT11}
\end{eqnarray}

Given a mode with $\lambda_0=0$ (the determinant in the $(\hat{\chi}_2,\hat{\psi}_1)$ sector of $M$ being zero), then just as noted in \cite{Alexandre2018}, the presence of a massless Goldstone boson is apparent, and the Goldstone theorem is thus  seen to hold when a non-Hermitian Hamiltonian has an antilinear symmetry.\footnote{While the $(\tilde{\chi}_2,\tilde{\psi}_1)$ sector of the mass matrix is not Hermitian, its antilinear symmetry cannot be realized in the complex conjugate pair realization because by being zero the Goldstone boson eigenvalue $\lambda_0$ is real. Consequently, $\lambda_1$ must be real too.} If we restrict the sign of the factor in the square root in $\lambda_{\pm}$ to be positive (the case considered in \cite{Alexandre2018}), then all mass eigenvalues are real. However, we note that we obtain a mode with $\lambda _0=0$ regardless of  the magnitude of this factor, and thus even obtain a Goldstone boson when the factor in the square root term is negative and mass eigenvalues appear in complex conjugate pairs. Moreover, as we show in Sec. \ref{S3} below, when the factor in the square root term is zero, in the $(\hat{\chi}_1,\hat{\psi}_2)$ sector the matrix $M$ becomes Jordan block.  The Goldstone boson mode is thus present in all three of the eigenvalue realizations that are allowed by antilinearity (viz. antilinear symmetry). Moreover, technically we do not even need to ascertain what the antilinear symmetry might even be, since as shown in \cite{Mannheim2018,Bender2010}, once we obtain an eigenvalue spectrum of the form that we have obtained in the   $(\hat{\chi}_1,\hat{\psi}_2)$ sector, the mass matrix must admit of an antilinear symmetry. Thus antilinearity implies this particular form for the mass spectrum, and this particular form for the mass spectrum implies antilinearity. Finally, we note that if in the $(\hat{\chi}_2,\hat{\psi}_1)$ sector we set $\mu^4=m_2^4$, then not only does $\lambda_1$ become zero just like $\lambda_0$, but as we show in Sec. \ref{S3} the entire sector becomes Jordan block, with the Goldstone boson eigenfunction itself then having the zero norm that is characteristic of Jordan-block systems.

\section{Eigenvectors of the Mass Matrix}
\label{S3}

To discuss the eigenvector spectrum of the mass matrix $M$, it is convenient to introduce the $PT$ theory $V$ operator. Specifically,  it was noted in  \cite{Mostafazadeh2002,Solombrino2002,Mannheim2018} that if a time-independent Hamiltonian has an antilinear symmetry there will always exist a time-independent operator $V$ that obeys the so-called pseudo-Hermiticity condition $VH=H^{\dagger}V$. If $V$ is invertible (this automatically being the case for any finite-dimensional matrix such as the mass matrix $M$ of interest to us here), then $H$ and $H^{\dagger}$ are isospectrally related according to $H^{\dagger}=VHV^{-1}$, to thus have the same set of eigenvalues. Since such an isospectral relation requires that the eigenvalues of $H$ be real or in complex pairs, pseudo-Hermiticity is equivalent to antilinearity.

If $H$ is not Hermitian, one has to introduce separate right- and left-Schr\"odinger equations in which $H$ acts to the right or to the left. Then from the relation $i\partial_t|n\rangle=H|n\rangle$ obeyed by solutions to the  right-Schr\"odinger equation we obtain $-i\partial_t\langle n|=\langle n|H^{\dagger}$, with $\langle n|$ then not being a solution to the left-Schr\"odinger equation as it does not obey $-i\partial_t\langle n|=\langle n|H$. Consequently in the non-Hermitian case the standard Dirac norm $\langle n(t)|n(t)\rangle=\langle n(0)|e^{iH^{\dagger}t}e^{-iHt}|n(0)\rangle$ is not time independent (i.e. not equal to $\langle n(0)|n(0)\rangle$), and one cannot use it as an inner product. However, the $V$ norm constructed from $V$ is time independent since 
\begin{eqnarray}
i\partial_t\langle n(t)|V|n(t)\rangle=\langle n(t)|(VH-H^{\dagger} V)|n(t)\rangle=0.
\label{GPT12}
\end{eqnarray}
Since we can set 
\begin{eqnarray}
-i\partial_t\langle n|=\langle n|H^{\dagger}=\langle n|VHV^{-1},\quad - i\partial_t\langle n|V=\langle n|VH,
\label{GPT13}
\end{eqnarray}
we see that it is the state $\langle n|V$ that is a solution to the left-Schr\"odinger equation and not the bra $\langle n|$ itself. Moreover, from (\ref{GPT13}) we obtain 
\begin{eqnarray}
\langle n(t)|V|n(t)\rangle=\langle n(0)|Ve^{iHt}e^{-iHt}|n(0)\rangle=\langle n(0)|V|n(0)\rangle,
\label{GPT14}
\end{eqnarray}
to thus confirm the time independence of the $V$ norm. Through the $V$ operator then we see that time independence of inner products and antilinear symmetry are equivalent. Given that $\langle L_n|=\langle n|V$ is a solution to the left-Schr\"odinger equation, in the event that it is also a left-eigenvector of $H$ and $|R_n\rangle$  is a right-eigenvector of $H$, in  the antilinear case the completeness relation is given not by $\sum |n\rangle\langle n|=I$ but by 
\begin{eqnarray}
\sum |n\rangle\langle n|V=\sum |R_n\rangle\langle L_n|=I
\label{GPT15}
\end{eqnarray}
instead. As shown in \cite{Mannheim2018a}, when charge conjugation is separately conserved,  the left-right $\langle R_n|V|R_m \rangle $ $V$-norm is the same as the overlap of the right-eigenstate $|R_n\rangle$ with its $PT$ conjugate (like $PT$ conjugation Hermitian conjugation is also antilinear). And more generally,  the $V$-norm is the same as the overlap of a state with its $CPT$ conjugate  \cite{Mannheim2018}.

In the special case where all the eigenvalues of a Hamiltonian are real and the eigenspectrum is complete, the Hamiltonian must either already obey $H=H^{\dagger}$ or be transformable by a (non-unitary) similarity transformation $S$ into one that does according to $SHS^{-1}=H^{\prime}=H^{\prime \dagger}$. For the primed system one has right-eigenvectors that obey 
\begin{eqnarray}
 i\partial_t|R_n^{\prime}\rangle=H^{\prime}|R_n^{\prime}\rangle,\quad -i\partial_t\langle R_n^{\prime}|=\langle R_n^{\prime}|H^{\prime},
\label{GPT16}
\end{eqnarray}
with the eigenstates of $H$ and $H^{\prime}$ being related by
\begin{eqnarray}
 |R_n^{\prime}\rangle=S|R_n\rangle,~~~\langle R_n^{\prime}|=\langle R_n|S^{\dagger}.
\label{GPT17}
\end{eqnarray}
On normalizing the eigenstates of the Hermitian $H^{\prime}$ to unity,  we obtain
\begin{eqnarray}
\langle R_n^{\prime} |R_m^{\prime}\rangle=\langle R_n|S^{\dagger}S|R_m\rangle=\delta_{m,n}.
\label{GPT18}
\end{eqnarray}
With $H^{\prime}=H^{\prime \dagger}$ we obtain
\begin{eqnarray}
SHS^{-1}=S^{\dagger -1}H^{\dagger}S^{\dagger},\quad S^{\dagger}SHS^{-1}S^{\dagger -1}=S^{\dagger}SH[S^{\dagger }S]^{-1}=H^{\dagger}.
\label{GPT19}
\end{eqnarray}
We can thus identify $S^{\dagger}S$ with $V$ when all energy eigenvalues are real and $H$ is diagonalizable, and as noted in \cite{Mannheim2018}, can thus establish that the $V$ norm is the $S^{\dagger}S$ norm, so that in this case $\langle L_n|R_m\rangle=\langle R_n|V|R_m\rangle=\langle R_n|S^{\dagger}S|R_m\rangle=\delta_{m,n}$ is positive definite.\footnote{As shown in \cite{Mannheim2018a}, to identify the $V$ norm with the $PT$ norm one has to choose the phase of the $PT$ conjugate of a state to be the same as the $PT$ eigenvalue of the state that is being conjugated. This prescription obviates any need to use the $PT$ theory $C$ operator norm that is described in \cite{Bender2007}, with the $PT$ norm then having the same positivity as the $V$ norm. Moreover, it was shown that not every $PT$ symmetric theory will possess a $PT$ theory $C$ operator, but all $PT$ theories will possess  $V$ and $PT$ norms.} The interpretation of the $V$ norms as probabilities is then secured, with their time independence ensuring that probability is preserved in time.

Having now presented the general non-Hermitian formalism, a formalism that holds in both wave mechanics and matrix mechanics \cite{Bender2007}, and holds in quantum field theory \cite{Mannheim2018}, we can apply it to the mass matrix $M$ given in (\ref{GPT9}). And while this matrix does arise in a quantum field theory,  all that matters in the following is that it has a non-Hermitian matrix structure. The matrix $M$ breaks up into two distinct two-dimensional blocks, and we can describe each of them by the generic 
\begin{eqnarray}
N=\pmatrix{C+A & iB \cr iB &C-A},
\label{GPT20}
\end{eqnarray}
where $A$, $B$ and $C$ are all real. The matrix $N$ is not Hermitian but does have a $PT$ symmetry if we set $P=\sigma_3$ and $T=K$ where $K$ effects complex conjugation. The eigenvalues of $N$ are given by 
\begin{eqnarray}
\Lambda_{\pm}=C\pm (A^2-B^2)^{1/2},
\label{GPT21}
\end{eqnarray}
and they are real if $A^2>B^2$ and in a complex conjugate pair if $A^2<B^2$, just as required of a non-Hermitian but $PT$-symmetric matrix. Additionally,  the relevant $S$ and $V$ operators are given by
\begin{eqnarray}
S&=&\frac{1}{2(A^2-B^2)^{1/4}}\pmatrix{(A+B)^{1/2}+(A-B)^{1/2}& i[(A+B)^{1/2}-(A-B)^{1/2}]\cr
 -i[(A+B)^{1/2}-(A-B)^{1/2}]&(A+B)^{1/2}+(A-B)^{1/2} },
 \nonumber\\
 S^{-1}&=&\frac{1}{2(A^2-B^2)^{1/4}}\pmatrix{(A+B)^{1/2}+(A-B)^{1/2}& -i[(A+B)^{1/2}-(A-B)^{1/2}]\cr
 i[(A+B)^{1/2}-(A-B)^{1/2}]&(A+B)^{1/2}+(A-B)^{1/2} },
 \nonumber\\
V&=&\frac{1}{(A^2-B^2)^{1/2}}\pmatrix{A & iB \cr -iB &A},\quad V^{-1}=\frac{1}{(A^2-B^2)^{1/2}}\pmatrix{A & -iB \cr iB &A},
\label{GPT22}
\end{eqnarray}
and they effect
\begin{eqnarray}
SNS^{-1}=N^{\prime}=\pmatrix{C+(A^2-B^2)^{1/2} & 0 \cr 0 &C-(A^2-B^2)^{1/2}},\quad VNV^{-1}=\pmatrix{C+A & -iB \cr -iB &C-A}=N^{\dagger}
\label{GPT23}
\end{eqnarray}
regardless of whether $A^2-B^2$ is positive or negative (if $A^2$ is less than $B^2$, then while not Hermitian $SNS^{-1}$ is still diagonal). However, as we elaborate on below,  we note that if $A^2-B^2$ is zero then $S$ and $V$ become undefined. Other than at $A^2-B^2=0$ the matrix $N^{\prime}=SNS^{-1}$ is diagonal, and with $N$ being given by $N=S^{-1}N^{\prime}S$, the right-eigenvectors of $N$ that obey $NR_{\pm}=\Lambda_{\pm}R_{\pm}$ are given by the columns of $S^{-1}$, and the left-eigenvectors of $N$ that obey $L_{\pm}N=\Lambda_{\pm}L_{\pm}$ are given by the rows of $S$. Given the right-eigenvectors one can also construct the left-eigenvectors by using the $V$ operator. When $A^2>B^2$ the left eigenvectors can be constructed as $\langle L_{\pm}|=\langle R_{\pm}|V$, and we obtain
\begin{eqnarray}
R_{+}&=&\frac{1}{2(A^2-B^2)^{1/4}}\pmatrix{(A+B)^{1/2}+(A-B)^{1/2}\cr
 i[(A+B)^{1/2}-(A-B)^{1/2}]}
 \nonumber\\
R_{-}&=&\frac{1}{2(A^2-B^2)^{1/4}}\pmatrix{ -i[(A+B)^{1/2}-(A-B)^{1/2}]\cr
 (A+B)^{1/2}+(A-B)^{1/2} }
\nonumber\\
L_{+}&=&\frac{1}{2(A^2-B^2)^{1/4}}\pmatrix{(A+B)^{1/2}+(A-B)^{1/2},& i[(A+B)^{1/2}-(A-B)^{1/2}]}
\nonumber\\
L_{-}&=& \frac{1}{2(A^2-B^2)^{1/4}}\pmatrix{
 -i[(A+B)^{1/2}-(A-B)^{1/2}],&(A+B)^{1/2}+(A-B)^{1/2} },
\label{GPT24}
\end{eqnarray}
and these eigenvectors are normalized according to the positive definite $\langle L_n|R_m\rangle=\langle R_n|V|R_m\rangle=\delta_{m,n}$, i.e. according to $L_{\pm}R_{\pm}=1$, $L_{\mp}R_{\pm}=0$. In addition $N$ and the identity $I$ can be reconstructed as
\begin{eqnarray}
N=|R_+\rangle \Lambda_+\langle L_+|+|R_-\rangle \Lambda_-\langle L_-|,\quad
I=|R_+\rangle \langle L_+|+|R_-\rangle\langle L_-|,
\label{GPT25}
\end{eqnarray}
to thus be diagonalized in the left-right basis. 

When $A^2-B^2$ is negative, the quantity $(A-B)^{1/2}$ is pure imaginary, and since $\langle R|$ is the Hermitian conjugate of $|R\rangle$, in the $A^2<B^2$ sector up to a phase we have $\langle L_{\mp}|=\pm \langle R_{\pm}|V$. If we set $A^2-B^2=-D^2$ where $D$ is real, the eigenvalues are $\Lambda_{\pm}=C\pm iD$. In a quantum theory with the mass matrix serving as the Hamiltonian, $|R_{\pm}\rangle$ would evolve as $e^{-i(C\pm iD)t}=e^{-iCt\pm Dt}$, while $\langle L_{\pm}|$ would evolve as $\langle R_{\mp}|V$, i.e. as $e^{iCt\mp Dt}$. As had been noted in general in \cite{Mannheim2018} and as found here, the only overlaps that would be non-zero would be  $\mp\langle L_{\pm}|R_{\pm}\rangle= \langle R_{\mp}|V|R_{\pm}\rangle=\pm i$, and they would be time independent. Since $\langle L_{\pm}|\neq \langle R_{\pm}|V$, these matrix elements would be transition matrix elements between  growing and decaying states. Such transition matrix elements are not required to be positive or to even be real.

While all  of these eigenstates and the $S$ and $V$ operators are well-defined as long as $A^2$ is not equal to $B^2$, at $A^2=B^2$ they all become singular. 
Moreover at $A^2=B^2$ the vectors $R_+$ and $R_-$ become identical to each other (i.e. equal up to an irrelevant overall phase), and equally $L_+$ and $L_-$ become identical too. The matrix $N$ thus loses both a left-eigenvector and a right-eigenvector at $A^2=B^2$ to then only have one left-eigenvector and only one right-eigenvector. At $A^2=B^2$ the two eigenvalues become equal ($\Lambda_+=\Lambda_-=C$) and have to share the same left- and right-eigenvectors.  The fact that $S$ becomes singular at $A^2=B^2$ means that $N$ cannot be diagonalized, with its eigenspectrum being incomplete. $N$ thus becomes a Jordan-block matrix that cannot be diagonalized.\footnote{Even though one loses diagonalizability when $A^2=B^2$, the matrix $N$ remains PT symmetric at $A^2=B^2$, as it is invariant under the $PT=\sigma_3K$ transformation for all values of its parameters as long as they are real.} Even though all of $L_{\pm}$ , $R_{\pm}$ become singular at $A^2=B^2$, $N$ still has left- and right-eigenvectors $L$ and $R$ that are given up to an arbitrary normalization by 
\begin{eqnarray}
L=\pmatrix{1 & i},\quad R=\pmatrix{1\cr i}, \quad LN=CN,\quad NR=CR,
\label{GPT26}
\end{eqnarray}
and no matter what that normalization might be, they obey the zero norm condition characteristic of Jordan-block matrices:
\begin{eqnarray}
LR=\pmatrix{1 & i}\pmatrix{1\cr i}=0.
\label{GPT27}
\end{eqnarray}
Even though the eigenspectrum of $N$ is incomplete, the vector space on which it acts is still complete. One can take the extra states to be 
\begin{eqnarray}
L^{\prime}=\pmatrix{1 & -i},\quad R^{\prime}=\pmatrix{1\cr -i},
\label{GPT28}
\end{eqnarray}
with $L^{\prime}R^{\prime}=0$, so that $R$ and $R^{\prime}$ span the space on which $N$ acts to the right, while $L$ and $L^{\prime}$ span the space on which $N$ acts to the left.

Comparing now with (\ref{GPT9}), we see that for the $(\hat{\chi}_1,\hat{\psi}_2)$  sector we have 
\begin{eqnarray}
C=\frac{2m_1^2m_2^2+m_2^4-3\mu^4}{2m_2^2},\quad A=\frac{2m_1^2m_2^2-3\mu^4-m_2^4}{2m_2^2},\quad B=\mu^2,
\label{GPT29}
\end{eqnarray}
while for the $(\hat{\chi}_2,\hat{\psi}_1)$  sector we have 
\begin{eqnarray}
C=\frac{m_2^4-\mu^4}{2m_2^2},\quad A=-\frac{(\mu^4+m_2^4)}{2m_2^2},\quad B=-\mu^2.
\label{GPT30}
\end{eqnarray}
From (\ref{GPT29}) and (\ref{GPT30}) the eigenvalues given in (\ref{GPT11}) follow. For the $(\hat{\chi}_1,\hat{\psi}_2)$ sector we thus have two eigenvectors with real eigenvalues if 
$(2m_1^2m_2^2-m_2^4-3\mu^4)^2>4\mu^4m_2^4$, two eigenvectors with complex conjugate  eigenvalues if  $(2m_1^2m_2^2-m_2^4-3\mu^4)^2<4\mu^4m_2^4$, and lose an eigenvector if 
$(2m_1^2m_2^2-m_2^4-3\mu^4)^2=4\mu^4m_2^4$. Since none of this affects the $(\hat{\chi}_2,\hat{\psi}_1)$ sector, for all three of the possible classes of eigenspectra associated with a non-Hermitian Hamiltonian with an antilinear symmetry we obtain a massless Goldstone boson.

For the $(\hat{\chi}_2,\hat{\psi}_1)$ sector the eigenvalues are $\lambda_0=0$ and $\lambda_1=m_2^2-\mu^4/m_2^2$. Both are them are real, and we shall take $m_2^4$ to not be less than $\mu^4$ so that $\lambda_1$ could not be negative. Additionally,  the left- and right-eigenvectors are given by
\begin{eqnarray}
L_0=\frac{1}{(m_2^4-\mu^4)^{1/2}}
\pmatrix{m_2^2, &i\mu^2},\quad R_0=\frac{1}{(m_2^4-\mu^4)^{1/2}}\pmatrix{m_2^2\cr i\mu^2}, 
\nonumber\\
L_1=\frac{1}{(m_2^4-\mu^4)^{1/2}}
\pmatrix{i\mu^2, & -m_2^2},\quad R_1=\frac{1}{(m_2^4-\mu^4)^{1/2}}\pmatrix{i\mu^2\cr -m_2^2},
\label{GPT31}
\end{eqnarray}
as normalized to 
\begin{eqnarray}
L_0R_0=1, \quad L_1 R_1=1, \quad L_0R_1=0,\quad L_1R_0=0.
\label{GPT32}
\end{eqnarray}
The Goldstone boson is thus properly normalized if one uses the left-right norm, with the two states in the $(\hat{\chi}_2,\hat{\psi}_1)$ sector forming a left-right orthonormal basis. Thus in the non-Hermitian case the standard Goldstone theorem associated with the spontaneous breakdown of a continuous symmetry continues to hold but the norm of the Goldstone boson has to be the positive left-right norm (or equivalently the $PT$ theory norm \cite{Alexandre2018}) rather than the standard positive Hermitian theory Dirac norm for which the theorem was first proved 
\cite{Nambu1960,Goldstone1961,Nambu1961,Goldstone1962}.

However, something unusual occurs if we set $\mu^2=m_2^2$. Specifically, the eigenvalue $\lambda_1$ becomes zero, to thus now be degenerate with $\lambda_0$. The eigenvectors $R_0$ and $R_1$ collapse onto a common single $R$ and $L_0$ and $L_1$ collapse onto a common single $L$, and the normalization coefficients given in (\ref{GPT31}) diverge. The  $(\hat{\chi}_2,\hat{\psi}_1)$ sector thus becomes of non-diagonalizable Jordan-block form. In this limit one can take the left- and right-eigenvectors to be 
\begin{eqnarray}
L=\pmatrix{1 & i},\quad R=\pmatrix{1\cr i}, 
\label{GPT33}
\end{eqnarray}
and they obey the zero norm condition 
\begin{eqnarray}
LR=0.
\label{GPT34}
\end{eqnarray}
As such this represents a new extension of the Goldstone theorem, and  even though the standard Goldstone theorem associated with the spontaneous breakdown of a continuous symmetry continues to hold, the norm of the Goldstone boson is now zero. Since a zero norm state can leave no imprint in a detector, we are essentially able to evade the existence of a massless Goldstone boson, in the sense that while it would still exist it would not be observable.

\section{Comparison with the work of Alexandre, Ellis, Millington and Seynaeve}
\label{S4}

If we do a functional variation of the action given in (\ref{GPT1}) we obtain 
\begin{eqnarray}
\delta I(\phi_1,\phi_2,\phi^*_1,\phi^*_2)&=&\int d^4x\bigg{[}[-\Box \phi_1+m_1^2\phi_1-\mu^2\phi_2-\frac{g}{2}\phi_1^2\phi_1^*]\delta \phi_1^*
+[-\Box \phi^*_1+m_1^2\phi^*_1+\mu^2\phi^*_2-\frac{g}{2}(\phi^*_1)^2\phi_1]\delta \phi_1
\nonumber\\
&+&[-\Box \phi_2-m_2^2\phi_2+\mu^2\phi_1]\delta \phi_2^*
+[-\Box \phi^*_2-m_2^2\phi^*_2-\mu^2\phi^*_1]\delta \phi_2
\nonumber\\
&+&\partial_{\mu}[\delta\phi^*_1\partial^{\mu}\phi_1+\delta\phi_1\partial^{\mu}\phi^*_1
+\delta\phi^*_2\partial^{\mu}\phi_2+\delta\phi_2\partial^{\mu}\phi^*_2]
\bigg{]}.
\label{GPT35}
\end{eqnarray}
With all variations held fixed at the surface, stationarity leads to
\begin{eqnarray}
&&-\Box \phi_1+m_1^2\phi_1-\mu^2\phi_2-\frac{g}{2}\phi_1^2\phi_1^*=0,
\nonumber\\
&&-\Box \phi_2-m_2^2\phi_2+\mu^2\phi_1=0,
\label{GPT36}
\end{eqnarray}
\begin{eqnarray}
&&-\Box \phi^*_1+m_1^2\phi^*_1+\mu^2\phi^*_2-\frac{g}{2}(\phi^*_1)^2\phi_1=0,
\nonumber\\
&&-\Box \phi^*_2-m_2^2\phi^*_2-\mu^2\phi^*_1=0,
\label{GPT37}
\end{eqnarray}
with these equations of motion being completely equivalent to (\ref{GPT7}). With these equations of motion one readily checks that the electric current $j_{\mu}=i(\phi^*_1\partial_{\mu} \phi_1-\phi_1\partial_{\mu} \phi^*_1)+i(\phi^*_2\partial_{\mu} \phi_2-\phi_2\partial_{\mu} \phi^*_2)$ given in (\ref{GPT4}) is conserved, just as it should be.

There is however an immediate problem with these equations of motion, namely if we complex conjugate (\ref{GPT36}) we obtain not (\ref{GPT37}) but 
\begin{eqnarray}
&&-\Box \phi^*_1+m_1^2\phi^*_1-\mu^2\phi^*_2-\frac{g}{2}(\phi^*_1)^2\phi_1=0,
\nonumber\\
&&-\Box \phi^*_2-m_2^2\phi^*_2+\mu^2\phi^*_1=0
\label{GPT38}
\end{eqnarray}
instead. The reason why this problem occurs is because while (\ref{GPT37}) is associated with $\partial I/\partial\phi_1$ and $\partial I/\partial\phi_2$, (\ref{GPT38}) is associated with $(\partial I/\partial\phi^*_1)^*=\partial I^*/\partial\phi_1$ and $(\partial I/\partial\phi^*_2)^*=\partial I^*/\partial\phi_2$ and $I$ is not equal to $I^*$ if $I$ is not Hermitian. A similar concern holds for (\ref{GPT7}) as not one of its four separate equations is left invariant under complex conjugation.

In order to get round this the authors of \cite{Alexandre2018} propose that  (\ref{GPT37}) not be valid, but rather one should use (\ref{GPT36}) and (\ref{GPT38}) instead. In order to achieve this the authors of \cite{Alexandre2018} propose that one add an additional surface term to (\ref{GPT35})  so that one no longer imposes stationarity with respect $\delta \phi_1$ and $\delta \phi_2$, but only  stationarity with respect to $\delta \phi^*_1$ and $\delta \phi^*_2$ alone.\footnote{The additional surface term is akin to the Hawking-Gibbons surface term used in general relativity. Specifically, in general relativity the variation of the Einstein-Hilbert action leads to variations of both $g_{\mu\nu}$ and its first derivative at the surface. Variations with respect to the derivatives are then cancelled by the Hawking-Gibbons term.} If one does use (\ref{GPT36}) and (\ref{GPT38}), the electric current $j_{\mu}$ is no longer conserved (i.e. the surface term that is to be introduced must carry off some electric charge), but instead it is the current 
\begin{eqnarray} 
j^{\prime}_{\mu}=i(\phi^*_1\partial_{\mu} \phi_1-\phi_1\partial_{\mu} \phi^*_1)- i(\phi^*_2\partial_{\mu} \phi_2-\phi_2\partial_{\mu} \phi^*_2)
\label{GPT39}
\end{eqnarray}
that is conserved in solutions to the equations of motion. As such, this $j^{\prime}_{\mu}$ current is a non-Noether current that is not associated with a symmetry of the action $I$ (unless the inclusion of the surface term then leads to one), and thus its spontaneous breakdown is somewhat different from the standard one envisaged in \cite{Nambu1960,Goldstone1961,Nambu1961,Goldstone1962}. Nonetheless, as noted in \cite{Alexandre2018}, when the scalar fields acquire vacuum expectation values, the mass matrix associated with (\ref{GPT36}) and (\ref{GPT38}) still has a zero eigenvalue. With the authors of \cite{Alexandre2018} showing that it is associated with the Ward identity for $j_{\mu}^{\prime}$, it can still be identified as a Goldstone boson. The work of  \cite{Alexandre2018}  thus breaks the standard connection between Goldstone bosons and symmetries of the action.

As such, the result of the authors of  \cite{Alexandre2018} is quite interesting as it provides possible new insight into the Goldstone theorem. However, the analysis somewhat obscures the issue  as it suggests that the generation of Goldstone bosons in non-Hermitian theories is quite different from the generation of Goldstone bosons in Hermitian theories. It is thus of interest to ask whether one could show that one could obtain Goldstone bosons in a procedure that is common to both Hermitian and non-Hermitian theories. To this end we need to find a way to exclude (\ref{GPT38}) and validate $(\ref{GPT37})$, as it is (\ref{GPT36}) and (\ref{GPT37}) that we used in our paper in  an approach that is completely conventional, one in which the surface term in (\ref{GPT35}) vanishes in the standard variational procedure way.

To reconcile (\ref{GPT36}) and (\ref{GPT37}) or to reconcile the equations of motion in (\ref{GPT7}) with complex conjugation it is instructive to make a particular similarity transformation on the fields, even though doing so initially appears to lead to another puzzle, the Hermiticity puzzle, which we discuss and resolve below. It is more convenient to seek a reconciliation for (\ref{GPT7}) first, so from $I(\chi_1,\chi_2,\psi_1,\psi_2)$ we identify canonical conjugates for $\phi_1$ and $\phi_2$ of the form $\Pi_1=\partial_t\psi_1$, $\Pi_2=\partial_t\psi_2$. With these conjugates  we introduce  \cite{Mannheim2018}
\begin{eqnarray}
S(\psi_1)=\exp\left[\frac{\pi}{2}\int d^3x \Pi_1(\textbf{x},t)\psi_1(\textbf{x},t)\right],\quad 
S(\psi_2)=\exp\left[\frac{\pi}{2}\int d^3x \Pi_2(\textbf{x},t)\psi_2(\textbf{x},t)\right],
\label{GPT40}
\end{eqnarray}
and obtain 
\begin{eqnarray}
S(\psi_1)\psi_1S^{-1}(\psi_1)=-i\psi_1,\quad S(\psi_1)\Pi_1S^{-1}(\psi_1)=i\Pi_1,\quad
S(\psi_2)\psi_2S^{-1}(\psi_2)=-i\psi_2,\quad S(\psi_2)\Pi_2S^{-1}(\psi_2)=i\Pi_2.
\label{GPT41}
\end{eqnarray}
Since these transformations preserve the equal-time commutation relations $[\psi_1(\textbf{x},t),\Pi_1(\textbf{y},t)]=i\delta^3(\textbf{x}-\textbf{y})$, $[\psi_2(\textbf{x},t),\Pi_2(\textbf{y},t)]=i\delta^3(\textbf{x}-\textbf{y})$, they are fully permissible transformations that do not modify the content of the field theory. Applying (\ref{GPT41}) to $I(\chi_1,\chi_2,\psi_1,\psi_2)$ we obtain 
\begin{eqnarray}
S(\psi_1)S(\psi_2)I(\chi_1,\chi_2,\psi_1,\psi_2)S^{-1}(\psi_2)S^{-1}(\psi_1)=I^{\prime}(\chi_1,\chi_2,\psi_1,\psi_2)
\label{GPT42}
\end{eqnarray}
where
\begin{eqnarray}
I^{\prime}(\chi_1,\chi_2,\psi_1,\psi_2)&=&\int d^4x \bigg{[}\frac{1}{2}\partial_{\mu}\chi_1\partial^{\mu}\chi_1+
\frac{1}{2}\partial_{\mu}\chi_2\partial^{\mu}\chi_2-
\frac{1}{2}\partial_{\mu}\psi_1\partial^{\mu}\psi_1-
\frac{1}{2}\partial_{\mu}\psi_2\partial^{\mu}\psi_2+
\frac{1}{2}m_1^2(\chi_1^2+\chi_2^2)
\nonumber\\
&+&
\frac{1}{2}m_2^2(\psi_1^2+\psi_2^2)-
\mu^2(\chi_1\psi_2-\chi_2\psi_1)
-\frac{g}{16}(\chi_1^2+\chi_2^2)^2
\bigg{]}.
\label{GPT43}
\end{eqnarray}
Stationary variation with respect to $\chi_1$, $\chi_2$, $\psi_1$, and $\psi_2$ replaces (\ref{GPT7}) by
\begin{eqnarray}
-\Box \chi_1&=&-m_1^2\chi_1+\mu^2\psi_2+\frac{g}{4}(\chi_1^3+\chi_1\chi_2^2),
\nonumber\\
-\Box \chi_2&=&-m_1^2\chi_2-\mu^2\psi_1+\frac{g}{4}(\chi_2^3+\chi_2\chi_1^2),
\nonumber\\
-\Box \psi_1&=&m_2^2\psi_1+\mu^2\chi_2,
\nonumber\\
-\Box \psi_2&=&m_2^2\psi_2-\mu^2\chi_1,
\label{GPT44}
\end{eqnarray}
and now each one of the equations of motion is separately invariant under  complex conjugation.

Returning now to the original $\phi_2$, $\phi^*_2$ fields we obtain 
\begin{eqnarray}
S(\psi_1)S(\psi_2)\phi_2S^{-1}(\psi_2)S^{-1}(\psi_1)=-i\phi_2,\quad S(\psi_1)S(\psi_2)\phi^*_2S^{-1}(\psi_2)S^{-1}(\psi_1)=-i\phi^*_2,
\label{GPT45}
\end{eqnarray}
so that $I(\phi_1,\phi_2,\phi^*_1,\phi^*_2)$ transforms into 
\begin{eqnarray}
I^{\prime}(\phi_1,\phi_2,\phi^*_1,\phi^*_2)=\int d^4x\left[\partial_{\mu}\phi^*_1\partial^{\mu}\phi_1-\partial_{\mu}\phi^*_2\partial^{\mu}\phi_2
+m_1^2\phi_1^*\phi_1+m_2^2\phi^*_2\phi_2+i\mu^2(\phi^*_1\phi_2-\phi^*_2\phi_1)-\frac{g}{4}(\phi^*_1\phi_1)^2\right],
\label{GPT46}
\end{eqnarray}
while the equations of motion become
\begin{eqnarray}
&&-\Box \phi_1+m_1^2\phi_1+i\mu^2\phi_2-\frac{g}{2}\phi_1^2\phi_1^*=0,
\nonumber\\
&&-\Box \phi_2-m_2^2\phi_2+i\mu^2\phi_1=0,
\label{GPT47}
\end{eqnarray}
\begin{eqnarray}
&&-\Box \phi^*_1+m_1^2\phi^*_1-i\mu^2\phi^*_2-\frac{g}{2}(\phi^*_1)^2\phi_1=0,
\nonumber\\
&&-\Box \phi^*_2-m_2^2\phi^*_2-i\mu^2\phi^*_1=0,
\label{GPT48}
\end{eqnarray}
and now there is no complex conjugation problem, with (\ref{GPT48}) being the complex conjugate of (\ref{GPT47}).\footnote{The appearance of a negative kinetic energy term for $\phi_2$ in (\ref{GPT46}) is only an artifact of the similarity transformation, since there are no such negative kinetic energy terms in  our starting $I(\chi_1,\chi_2,\psi_1,\psi_2)$ and one cannot change the signature of a Hilbert space by a similarity transformation.} In addition we note under the transformations given in (\ref{GPT45}) the equations given in (\ref{GPT38}) transform into 
\begin{eqnarray}
&&-\Box \phi^*_1+m_1^2\phi^*_1+i\mu^2\phi^*_2-\frac{g}{2}(\phi^*_1)^2\phi_1=0,
\nonumber\\
&&-\Box \phi^*_2-m_2^2\phi^*_2+i\mu^2\phi^*_1=0.
\label{GPT49}
\end{eqnarray}
If we now switch the sign of $\phi_2^*$, (\ref{GPT47}) is unaffected, while (\ref{GPT49}) becomes
\begin{eqnarray}
&&-\Box \phi^*_1+m_1^2\phi^*_1-i\mu^2\phi^*_2-\frac{g}{2}(\phi^*_1)^2\phi_1=0,
\nonumber\\
&&-\Box \phi^*_2-m_2^2\phi^*_2-i\mu^2\phi^*_1=0.
\label{GPT50}
\end{eqnarray}
We recognize (\ref{GPT50}) as being (\ref{GPT48}). With (\ref{GPT47}) being unaffected by the switch in sign of $\phi_2^*$, the mass matrix based on (\ref{GPT47}) and (\ref{GPT48}) is the same 
as the mass matrix based on (\ref{GPT47}) and (\ref{GPT50}). However, since all we have done in going from (\ref{GPT36}), (\ref{GPT37}) and (\ref{GPT38}) is make similarity transformations that leave determinants invariant, the eigenvalues associated with (\ref{GPT36}) and (\ref{GPT37}) (i.e. with (\ref{GPT9})) on the one hand and the eigenvalues associated with (\ref{GPT36}) and (\ref{GPT38}) on the other hand must be the same. And indeed this is exactly found to be the case, with all four of the eigenvalues given in \cite{Alexandre2018} being precisely the ones given in our (\ref{GPT11}). One can thus obtain the same mass spectrum as that obtained in \cite{Alexandre2018} using a completely conventional variational procedure. 

In addition, we note that with (\ref{GPT47}) and (\ref{GPT50}) the current $j^{\prime}_{\mu}$ given in (\ref{GPT39}) that is used in \cite{Alexandre2018} now is conserved. In fact, under the transformations given in (\ref{GPT45}) the $j_{\mu}$ current given in (\ref{GPT4}) transforms into $j_{\mu}^{\prime}$. Thus all that is needed to bring the study of \cite{Alexandre2018}) into the conventional Goldstone framework (standard variation procedure, standard spontaneous breakdown of a symmetry of the action) is to first make a similarity transformation.

Now the reader will immediately object to what we have done since now the $\mu^2(\chi_1\psi_2-\chi_2\psi_1)$ term in (\ref{GPT43}) and the $i\mu^2(\phi^*_1\phi_2-\phi^*_2\phi_1)$ term in (\ref{GPT46}) are both invariant under complex conjugation. Then with the actions in (\ref{GPT43})  and (\ref{GPT46}) then seemingly being Hermitian, we are seemingly back to the standard Hermitian situation where the Goldstone theorem readily holds, and we have seemingly gained nothing new.  However, it cannot actually be the case that action in (\ref{GPT43}) could be Hermitian, since similarity transformations cannot change the eigenvalues of the mass matrix $M$ given in (\ref{GPT9}), and as we have seen for certain values of parameters the eigenvalues can be complex or $M$ could even be Jordan block. We thus need to explain how, despite its appearance, a seemingly Hermitian action might not actually be Hermitian. The answer to this puzzle has been provided in \cite{Mannheim2018}, and we describe it below.

However, before doing so we note that there are two other approaches that could also achieve a  reconciliation. The first alternative involves starting with  the fields $\chi_1$, $\chi_2$, $\psi_1$, $\psi_2$ as the fields that define the theory, and $I(\chi_1, \chi_2, \psi_1, \psi_2)$ as the input action. In this case one immediately obtains the equations of motion given in (\ref{GPT7}). As they stand these equations are inconsistent if all the four fields are Hermitian. If we take $\chi_1$ and $\chi_2$ to be Hermitian, then these equations force $\psi_1$ and $\psi_2$ to be anti-Hermitian. And if $\psi_1$ and $\psi_2$ are taken to be anti-Hermitian, both the equations of motion and the action given in (\ref{GPT7}) then are invariant under a complex  conjugation (i.e. Hermitian conjugation) in which $\psi_1$ and $\psi_2$ transform into $-\psi_1$ and $-\psi_2$. Moreover, in such a case the $-i\psi_1$ and $-i\psi_2$ fields that are generated through the similarity transformations given in (\ref{GPT41}) that would then be Hermitian. Of course then the interaction term given in (\ref{GPT6}) would be Hermitian as well, and we again have a seemingly Hermitian theory.

Now suppose we do take $\psi_1$ and $\psi_2$ to be anti-Hermitian. Then if we start with $I(\chi_1, \chi_2, \psi_1, \psi_2)$  we cannot get back to $I(\phi_1,\phi_2,\phi^*_1,\phi^*_2)$ given in (\ref{GPT1}), since in the correspondence given in (\ref{GPT5}) $\phi_2^*$ was recognized as the conjugate of a $\phi_2=\psi_1+i\psi_2$ field that was expanded in terms of Hermitian $\psi_1$ and $\psi_2$. If we now take $\phi_2$ to still be defined as $\phi_2=\psi_1+i\psi_2$, the associated $\phi_2^*$ would now be given by $-(\psi_1-i\psi_2)$, and thus equal to minus the previous $\psi_1-i\psi_2$ used in (\ref{GPT5}). With this definition a rewriting of $I(\chi_1, \chi_2, \psi_1, \psi_2)$ in the $(\phi_1,\phi_2,\phi^*_1,\phi^*_2)$ basis would yield 
\begin{eqnarray}
I(\phi_1,\phi_2,\phi^*_1,-\phi^*_2)=\int d^4x\left[\partial_{\mu}\phi^*_1\partial^{\mu}\phi_1-\partial_{\mu}\phi^*_2\partial^{\mu}\phi_2
+m_1^2\phi_1^*\phi_1+m_2^2\phi^*_2\phi_2-\mu^2(\phi^*_1\phi_2+\phi^*_2\phi_1)-\frac{g}{4}(\phi^*_1\phi_1)^2\right],
\label{GPT51}
\end{eqnarray}
and equations of motion
\begin{eqnarray}
&&-\Box \phi_1+m_1^2\phi_1-\mu^2\phi_2-\frac{g}{2}\phi_1^2\phi_1^*=0,
\nonumber\\
&&-\Box \phi_2-m_2^2\phi_2+\mu^2\phi_1=0,
\label{GPT52}
\end{eqnarray}
\begin{eqnarray}
&&-\Box \phi^*_1+m_1^2\phi^*_1-\mu^2\phi^*_2-\frac{g}{2}(\phi^*_1)^2\phi_1=0,
\nonumber\\
&&-\Box \phi^*_2-m_2^2\phi^*_2+\mu^2\phi^*_1=0.
\label{GPT53}
\end{eqnarray}
Now complex conjugation can be consistently applied, with  (\ref{GPT53}) being derivable from (\ref{GPT50}) by complex conjugation. And again it is $j^{\prime}_{\mu}$ that is conserved.

A second alternative approach is to reinterpret the meaning of the star operator used in $\phi_1^*$ and $\phi_2^*$. Instead of taking it to denote Hermitian conjugation, we could instead take it denote $CPT$ conjugation, i.e. $\phi_1^*=CPT\phi_1TPC$, $\phi_2^*=CPT\phi_2TPC$. Now we had noted in (\ref{GPT2}) that in order to enforce $CPT$ symmetry on $I(\phi_1,\phi_2,\phi^*_1,\phi^*_2)$  we took $\phi_1$ to be even and $\phi_2$ to be odd under $CPT$, and we had noted that in general a scalar field should be $CPT$ even (i.e. the same  $CPT$ parity as the $CPT$ even fermionic $\bar{\psi}\psi$ \cite{Mannheim2018}). However, if we apply  the similarity transformation given in (\ref{GPT41}) to $\phi_2=\psi_1+i\psi_2$ to get $-i\phi_2$, that would change the $CPT$ parity. Thus while $\phi_2$ has negative $CPT$ parity it is similarity equivalent to a field that has the conventional positive $CPT$ parity, with the transformed $I^{\prime}(\phi_1,\phi_2,\phi^*_1,\phi^*_2)$ and the resulting equations of motion now being $CPT$ symmetric if $\phi_2$ is taken to have positive $CPT$ parity, viz. $CPT\phi_2TPC=\phi_2^*$,  $CPT\phi^*_2TPC=\phi_2$. (We leave $\phi_1$ as given in (\ref{GPT2}), viz. $CPT\phi_1TPC=\phi^*_1$,  $CPT\phi^*_1TPC=\phi_1$.) 

The difficulty identified by the authors of \cite{Alexandre2018} can thus be resolved by a judicious choice of which fields are Hermitian and which are anti-Hermitian, by  a judicious choice of which fields are $CPT$ even and which are $CPT$ odd, or by similarity transformations that generate complex phases that affect both Hermiticity and $CPT$ parity. However in all of these such resolutions we are led to theories that now appear to be Hermitian and yet for certain values of parameters could not be, and so we need to address this issue.

\section{Resolution of the Hermiticity Puzzle}
\label{S5}

In \cite{Mannheim2018} the issues of the generality of $CPT$ symmetry and the nature of Hermiticity were addressed. In regard to Hermiticity it was shown that Hamiltonians that appear to be Hermitian need not be,
since Hermiticity or self-adjointness is determined not by superficial inspection of the appearance of the Hamiltonian but by construction of asymptotic boundary conditions, as they determine whether or not one could drop surface terms in an integration by parts. And even if one could drop surface terms we still may not get Hermiticity because of the presence of factors of $i$ in $H$ that could affect complex conjugation.  In regard to $CPT$ it was shown that if one imposes only two requirements, namely the time independence of  inner products and invariance under the complex Lorentz group, it follows that the Hamiltonian must have an antilinear $CPT$ symmetry. Since this analysis involves no Hermiticity requirement, the $CPT$ theorem is thus extended to the non-Hermitian case. As noted above, the time independence of inner products is achieved if the theory has any antilinear symmetry with the left-right $V$ norm being the inner product one has to use. Complex Lorentz invariance then forces the antilinear symmetry to be $CPT$.

In field theories one ordinarily constructs actions so that they are invariant under the real Lorentz group. However, the same analysis that shows that actions with spin zero Lagrangians are invariant under the real Lorentz group (the restricted Lorentz group)  also shows that they are invariant under the complex one (the proper Lorentz group that includes $PT$ transformations for coordinates and $CPT$ transformations for spinors). Specifically, the action $I=\int d^4x L(x)$ with spin zero $L(x)$ is left invariant under real Lorentz transformations of the form $\exp(iw^{\mu\nu}M_{\mu\nu})$ where the six antisymmetric $w^{\mu\nu}=-w^{\nu\mu}$ are real parameters and the six $M_{\mu\nu}=-M_{\nu\mu}$ are the generators of the Lorentz group. To see this we note that with  $M_{\mu\nu}$ acting on the Lorentz spin zero  $L(x)$ as $x_{\mu}p_{\nu}-x_{\nu}p_{\mu}$, under an infinitesimal Lorentz transformation the change in the action is  given by $\delta I=2w^{\mu\nu}\int d^4x x_{\mu}\partial_{\nu}L(x)=2w^{\mu\nu}\int d^4x [\partial_{\nu}[x_{\mu}L(x)]-\eta_{\mu\nu}L(x)]$, and since the metric $\eta_{\mu\nu}$ is symmetric and $w_{\mu\nu}$ is antisymmetric, thus given by $\delta I=2w^{\mu\nu}\int d^4x \partial_{\nu}[x_{\mu}L(x)]$. Since the change in the action is a total divergence, the familiar invariance of the action under real Lorentz transformations is secured. However, we note that nothing in this argument depended on $w^{\mu\nu}$ being real, with the change in the action still being a total divergence even if $w^{\mu\nu}$ is complex. The action $I=\int d^4x L(x)$ is thus actually invariant under complex Lorentz transformations as well and not just  under real ones, with complex Lorentz invariance thus being  just as natural to physics as real Lorentz invariance. 

For our purposes here we note that the Lorentz invariant scalar field action $I(\phi_1,\phi_2,\phi^*_1,\phi^*_2)$ given in (\ref{GPT1}) is thus invariant not just under real Lorentz transformations but under complex ones as well. Since in the above we constructed a time-independent inner product for this theory, the $I(\phi_1,\phi_2,\phi^*_1,\phi^*_2)$ action thus must have $CPT$ symmetry. And indeed we explicitly showed in (\ref{GPT2}) that this was in fact the case.

Since theories can thus be $CPT$ symmetric without needing to be Hermitian, it initially looks as though the two concepts are distinct. However, the issue of Hermiticity was addressed in \cite{Mannheim2018}, and the unexpected outcome of that study was that the only allowed Hamiltonians that one could construct that were $CPT$ invariant would have exactly the same structure as (or be similarity equivalent to) the ones one constructs in Hermitian theories, namely presumed Hermitian combinations of fields and all coefficients real.\footnote{While for instance $I(\chi_1,\chi_2,\psi_1,\psi_2)$ of (\ref{GPT6}) contains factors of $i$, its similarity transformed $I^{\prime}(\chi_1,\chi_2,\psi_1,\psi_2)$ given in (\ref{GPT43}) does not. Moreover this is even true of the $H=p^2+ix^3$ paradigm for $PT$ symmetry. With $S(\theta)=\exp(-\theta px)$ effecting the $[x,p]=i$ preserving $S(\theta)pS(-\theta)=\exp(-i\theta)p$, $S(\theta)xS(-\theta)=\exp(i\theta)x$, transforming with $S(\pi/2)$ effects $S(\pi/2)(p^2+ix^3)S(-\pi/2)=-p^2+x^3$, and in passing we note that $S(\pi)$ effects $S(\pi)(p^2+ix^3)S(-\pi)=p^2-ix^3=(p^2+ix^3)^{\dagger}$. In fact in \cite{Mannheim2018} it was shown in general that $CPT$ invariance of a relativistic theory entails that one can always find an appropriate similarity transformation that would bring the Hamiltonian  to a form in which all coefficients are real.} These are precisely the theories that one ordinarily refers to as Hermitian. However, thus turns out to not necessarily be the case since theories can appear to be Hermitian but not actually be so.

To illustrate the above remarks it is instructive to consider some explicit examples, one involving behavior in time and the other involving behavior in space. For behavior in time consider the neutral scalar field with action $I_{\rm S}=\int d^4x [\partial_{\mu}\phi\partial^{\mu}\phi-m^2\phi^2]/2$ and Hamiltonian $H=\int d^3x[\dot{\phi}^2+\vec{\nabla}\phi\cdot \vec{\nabla}\phi+m^2\phi^2]/2$. Solutions to the  wave equation $-\ddot{\phi}+\nabla^2\phi-m^2\phi=0$  obey $\omega^2(\textbf{k})=\textbf{k}^2+m^2$. Thus the poles in the scalar field propagator are at $\omega(\textbf{k})=\pm[\textbf{k}^2+m^2]^{1/2}$, the field can be expanded as $\phi(\textbf{x},t)=\sum [a(\textbf{k})\exp(-i\omega(\textbf{k}) t+i\textbf{k}\cdot\textbf{x})+a^{\dagger}(\textbf{k})\exp(+i\omega(\textbf{k}) t-i\textbf{k}\cdot\textbf{x})]$,  and the Hamiltonian is given by $H=\sum [\textbf{k}^2+m^2]^{1/2}[a^{\dagger}(\textbf{k})a(\textbf{k})+a(\textbf{k})a^{\dagger}(\textbf{k})]/2$.  

For either sign of $m^2$ the $I_{\rm S}$ action is CPT symmetric, and for both signs $I_{\rm S}$ appears to be Hermitian. For $m^2>0$, $H$ and $\phi(\textbf{x},t)$ are indeed Hermitian and all frequencies are real. However, for $m^2<0$, frequencies become complex when $\textbf{k}^2<-m^2$. The poles in the propagator move into the complex plane, the field $\phi(\textbf{x},t)$ then contains modes that grow or decay exponentially in time,\footnote{Since the action is $CPT$ symmetric, if there are to be any complex frequencies they must appear in complex conjugate pairs.} while $H$ contains energies that are complex. Thus neither $H$ nor  $\phi$ is Hermitian even though $I_{\rm S}$ appears to be so. 

For behavior in space consider the Pais-Uhlenbeck two-oscillator theory \cite{Pais1950} as studied in \cite{Bender2008a,Bender2008b}. In the theory there are  two sets of oscillator operators, which  obey $[z,p_z]=i$, $[x,p_x]=i$, and the  Hamiltonian is given by
\begin{eqnarray}
H_{\rm PU}(\omega_1,\omega_2)=\frac{p_x^2}{2\gamma}+p_zx+\frac{\gamma}{2}\left(\omega_1^2+\omega_2^2 \right)x^2-\frac{\gamma}{2}\omega_1^2\omega_2^2z^2.
\label{GPT54}
\end{eqnarray}
As noted in \cite{Bender2008a} this theory is $PT$ symmetric,  and as noted in \cite{Bender2008b} it in addition is the non-relativistic limit  of a relativistic fourth-order neutral scalar field theory, one whose $CPT$ symmetry reduces to $PT$ symmetry in the non-relativistic limit. Initially the $\omega_1$ and $\omega_2$ frequencies are taken to be real and positive, and the energy eigenvalues are the real and positive $E(n_1,n_2)=(n_1+1/2)\omega_1+(n_2+1/2)\omega_2$.

However, if we now take the two frequencies to be equal to $\omega$,  the Hamiltonian takes the form 
\begin{eqnarray}
H_{\rm PU}(\omega)=\frac{p^2}{2\gamma}+p_zx+\gamma\omega^2x^2-\frac{\gamma}{2}\omega^4z^2,
\label{GPT55}
\end{eqnarray}
and while $H_{\rm PU}(\omega)$ looks to be just as Hermitian as before, the Hamiltonian turns out to be Jordan block \cite{Mannheim2005,Bender2008b}, to thus necessarily not be Hermitian at all. Since the $CPT$ invariance of $H_{\rm PU}(\omega_1,\omega_2)$ is not affected by setting $\omega_1=\omega_2$, $H_{\rm PU}(\omega)$ is $CPT$ symmetric.

Moreover, if we take the two frequencies to be in a complex pair $\omega_1=\alpha +i\beta$, $\omega_2=\alpha -i\beta$ with $\alpha>0$, $\beta>0$, the Hamiltonian takes the form \cite{Mannheim2018}
\begin{eqnarray}
H_{\rm PU}(\alpha,\beta)=\frac{p^2}{2\gamma}+p_zx+\gamma(\alpha^2-\beta^2)x^2-\frac{\gamma}{2}(\alpha^2+\beta^2)^2z^2.
\label{GPT56}
\end{eqnarray}
The $H_{\rm PU}(\alpha,\beta)$ Hamiltonian still looks to be Hermitian but its energy eigenvalues are now in complex conjugate pairs. With all the coefficients in  $H_{\rm PU}(\alpha,\beta)$ being real, $H_{\rm PU}(\alpha,\beta)$ is $CPT$ symmetric. Thus all three of $H_{\rm PU}(\omega_1,\omega_2)$, $H_{\rm PU}(\omega)$ and $H_{\rm PU}(\alpha,\beta)$ are $CPT$ invariant and for all three of them all coefficients are real, just as required by \cite{Mannheim2018}. However, despite their appearance, $H_{\rm PU}(\omega)$ and $H_{\rm PU}(\alpha,\beta)$ are necessarily non-Hermitian.

As written in (\ref{GPT54}), $H_{\rm PU}(\omega_1,\omega_2)$ is actually not Hermitian (or self-adjoint) either \cite{Bender2008a}, since the real issue is not the appearance of the Hamiltonian but whether in an integration by parts one can drop spatially asymptotic surface terms. To see this we make a standard representation of the momentum operators of the form $p_z=-i\partial_z$, $p_x=-i\partial_x$, and find that for the Schr\"odinger problem associated with $H_{\rm PU}(\omega_1,\omega_2)$ the ground state wave function $\psi_0(z,x)$ with energy $E(0,0)=(\omega_1+\omega_2)/2$ is given by
\begin{eqnarray}
\psi_0(z,x)={\rm exp}\left[\frac{\gamma}{2}(\omega_1+\omega_2)\omega_1\omega_2
z^2+i\gamma\omega_1\omega_2zx-\frac{\gamma}{2}(\omega_1+\omega_2)x^2\right].
\label{GPT57}
\end{eqnarray}
Since this wave function is divergent at large $z$ it is not normalizable (though it is convergent at large  $x$). Consequently, one cannot throw surface terms away in an integration by parts, and despite its appearance $H_{\rm PU}(\omega_1,\omega_2)$ is not self-adjoint. In the three realizations described above ($\omega_1$ and $\omega_2$ real and unequal, real and equal, in a complex conjugate pair) we find that $\omega_1+\omega_2$ and $\omega_1\omega_2$ are all real and positive. Thus in all three realizations the wave functions diverge at large $z$, and in all three cases the Hamiltonian is not self-adjoint when acting on $\psi_0(z,x)$

By the same token one cannot throw surface terms away for $p_z$ and $p_x$ when they act on the eigenstates of $H_{\rm PU}(\omega_1,\omega_2)$. Thus even though $p_z$ and $p_x$ are Hermitian when acting on their own eigenstates they are not Hermitian when acting on the eigenstates of $H_{\rm PU}(\omega_1,\omega_2)$. Thus building a Hamiltonian out of Hermitian operators (i.e. ones that are Hermitian when acting on their own eigenstates)  does not necessarily produce a Hamiltonian that is Hermitian when the Hamiltonian acts on its own eigenstates. In fact, until one has constructed the eigenstates of a Hamiltonian one cannot even tell whether or not a Hamiltonian is Hermitian at all. One thus cannot declare a Hamiltonian to be Hermitian just by superficial inspection. Rather, one has to construct its eigenstates first and look at their asymptotic behavior. 

In order to obtain eigenvectors for $H_{\rm PU}(\omega_1,\omega_2)$ that are normalizable the authors of \cite{Bender2008a} made the similarity transformation
\begin{eqnarray}
y=e^{\pi p_zz/2}ze^{-\pi p_zz/2}=-iz,\quad q=e^{\pi p_zz/2}p_ze^{-\pi p_zz/2}=
ip_z,
\label{GPT58}
\end{eqnarray}
on the operators of the theory so that $[y,q]= i$. Under this same transformation $H_{\rm PU}(\omega_1,\omega_2)$ transforms into
\begin{eqnarray}
e^{\pi p_zz/2}H_{\rm PU}(\omega_1,\omega_2)e^{-\pi p_zz/2}=\bar{H}_{PU}(\omega_1,\omega_2)=\frac{p^2}{2\gamma}-iqx+\frac{\gamma}{2}\left(\omega_1^2+\omega_2^2
\right)x^2+\frac{\gamma}{2}\omega_1^2\omega_2^2y^2,
\label{GPT59}
\end{eqnarray}
where for notational simplicity we have replaced $p_x$ by $p$, so that $[x,p]=i$. With the eigenvalue $z$ of the operator $z$ being replaced in $\psi_0(z,x)$ by $-iz$ (i.e. continued into the complex $z$ plane), the eigenfunctions are now normalizable.\footnote{As noted in \cite{Mannheim2013,Mannheim2018}, the analog statement for the Pais-Uhlenbeck two-oscillator theory path integral is that the path integral measure has to be continued into the complex plane in order to get the path integration to converge. A similar situation pertains to the path integral associated with the relativistic neutral scalar field theory with action $I_S=(1/2)\int d^4x[\partial_{\mu}\partial_{\nu}\phi\partial^{\mu}\partial^{\nu}\phi-(M_1^2+M_2^2)\partial_{\mu}\phi\partial^{\mu}
\phi+M_1^2M_2^2\phi^2]$, a theory whose non-relativistic limit is the Pais-Uhlenbeck theory.} When acting on the eigenfunctions of $\bar{H}_{PU}(\omega_1,\omega_2)$ the  $y$ and $q=-i\partial_y$ operators are Hermitian (as are $x$ and $p=-i\partial_x$). However, as the presence of the factor $i$ in the $-iqx$ term indicates, $\bar{H}_{PU}(\omega_1,\omega_2)$ is not Hermitian. Since in general to establish Hermiticity one has to integrate by parts, drop surface terms and complex conjugate, we see that while we now can drop surface terms for $\bar{H}_{PU}(\omega_1,\omega_2)$ we do not recover the generic $H_{ij}=H_{ji}^*$ when we complex conjugate, even as we can now drop surface terms for the momentum operators when they act on the eigenstates of $\bar{H}_{PU}(\omega_1,\omega_2)$ and achieve Hermiticity for them.\footnote{The use of the similarity transformations given in (\ref{GPT58}) parallels the use of (\ref{GPT40}) in Sec. \ref{S4}. However, while using the similarity transformation of (\ref{GPT40}) was mainly a convenience, for $H_{PU}(\omega_1,\omega_2)$ the similarity transformation of (\ref{GPT58}) is a necessity because of the need to construct normalizable wave functions. The presence of the factor $i$ in (\ref{GPT59}) is thus related to the intrinsic structure of the Pais-Uhlenbeck theory.}

When $\omega_1$ and $\omega_2$ are real and unequal, the eigenvalues of the Hamiltonian $\bar{H}_{PU}(\omega_1,\omega_2)$ are all and the eigenspectrum (two sets of harmonic oscillators) is complete. In that case $\bar{H}_{PU}(\omega_1,\omega_2)$ can actually be brought to a form in which it is Hermitian by a similarity transformation. Specifically, one introduces an operator $Q$
\begin{eqnarray}
Q=\alpha pq+\beta xy,\quad \alpha=\frac{1}{\gamma\omega_1\omega_2}{\rm log}\left(\frac{\omega_1+\omega_2}{\omega_1-\omega_2}\right),\quad \beta=\alpha\gamma^2\omega_1^2\omega_2^2,
\label{GPT60}
\end{eqnarray}
and obtains 
\begin{eqnarray}
&&\bar{H}_{PU}^{\prime}(\omega_1,\omega_2)=e^{-Q/2}\bar{H}_{PU}(\omega_1,\omega_2)e^{Q/2}=
\frac{p^2}{2\gamma}+\frac{q^2}{2\gamma\omega_1^2}+
\frac{\gamma}{2}\omega_1^2x^2+\frac{\gamma}{2}\omega_1^2\omega_2^2y^2.
\label{GPT61}
\end{eqnarray}
With the $Q$  similarity transformation not affecting the asymptotic behavior of the eigenstates of $\bar{H}_{PU}(\omega_1,\omega_2)$, and with $y$, $q$, $x$, and $p$ thus all being Hermitian when acting on the eigenstates of $\bar{H}_{PU}^{\prime}(\omega_1,\omega_2)$, the Hermiticity of $\bar{H}_{PU}^{\prime}(\omega_1,\omega_2)$ in the conventional Dirac sense is established. We can thus regard $\bar{H}_{PU}(\omega_1,\omega_2)$  with real and unequal $\omega_1$ and $\omega_2$ as being Hermitian in disguise. Moreover, in addition we note that since $Q$ becomes singular at $\omega_1=\omega_2$, at $\omega_1=\omega_2$ $\bar{H}_{PU}(\omega_1,\omega_2)$ cannot be diagonalized, to thus confirm that $H_{PU}(\omega)$  is  Jordan block.\footnote{The transformation with $Q$ is the analog of the transformation of the spontaneously broken scalar field theory mass matrix given in (\ref{GPT22}), and the singularity in $Q$ at $\omega_1=\omega_2$ is the analog of that in (\ref{GPT22}) when $A=B$.} In general then we see that a Hamiltonian may not be Hermitian even though it may appear to be so, and may be (similarity equivalent to) Hermitian even when it does not appear to be so. And moreover, one cannot tell beforehand, as one needs to first solve the theory and see what its solutions look like.

Other then possibly needing to continue into the complex plane in order to get convergence, when a Hamiltonian has all eigenvalues real and eigenspectrum complete it is always possible to similarity transform it into a form in which it is Hermitian in the standard Dirac sense. If a Hamiltonian obeys $H=H^{\dagger}$, then under a similarity transform that effects $H^{\prime}=SHS^{-1}$, we note that $H^{\prime \dagger}=S^{-1 \dagger}H^{\dagger}S^{\dagger}=S^{-1 \dagger}HS^{\dagger}=S^{-1 \dagger}S^{-1}H^{\prime}SS^{\dagger}=[SS^{\dagger}]^{-1}H^{\prime}SS^{\dagger}$. Thus unless $S$ is unitary $H^{\prime \dagger}$ is not equal to $H^{\prime}$, with the $H_{ij}=H_{ji}^*$ Hermiticity condition being a condition that is not preserved under a general similarity transformation. Thus if  one starts with some general $H^{\prime}$ that does not obey $H^{\prime}=H^{\prime \dagger}$, it might be similarity equivalent to a Hermitian $H$ but one does not know a priori. It only will be similarity equivalent to a Hermitian $H$ if the eigenvalues of $H^{\prime}$ are all real and the eigenspectrum is complete. And the necessary condition for that to be the case is that $H^{\prime}$ possess an antilinear symmetry. However, unlike a Hermiticity condition a commutation relation  is preserved under a similarity transformation (even a commutation relation that involves an antilinear operator \cite{Mannheim2018}), with antilinear operators being more versatile than Hermitian operators. So much so in fact that in \cite{Mannheim2018} it was argued that one should use $CPT$ symmetry as the guiding principle for constructing quantum theories rather than Hermiticity.\footnote{Thus rather than being optional,  according to  \cite{Mannheim2018} one has to interpret the star symbol in (\ref{GPT1}) as a $CPT$ transform.}

When we characterize an operator such as $z$, $p_z$, $x$, or $p_x$ as being Hermitian we are only referring to representations of the $[z,p_z]=i$ and $[x,p_x]=i$ commutation relations, without any reference to a Hamiltonian that might contain these operators. A Hamiltonian can thus be built out of Hermitian operators and can have all real coefficients, and yet not be Hermitian itself. The equal-frequency and complex-frequency Pais-Uhlenbeck models are particularly instructive in this regard. In the equal-frequency case none of  the $z$, $p_z$, $x$, or $p_x$ operators themselves are Jordan block, only $H_{PU}(\omega)$ is. The spectrum of eigenstates of the position and momentum operators are complete, and all are contained in the space on which $H_{\rm PU}(\omega)$ acts. However, not all of these states are eigenstates of the Hamiltonian \cite{Bender2008b}, with the one-particle sector of $H_{PU}(\omega)$ behaving just like the example given in (\ref{GPT26}) and (\ref{GPT28}). Moreover, in the complex $H_{\rm PU}(\alpha,\beta)$ case all the eigenvalues of the position and momentum operators are real even though those of the Hamiltonian that is built out of them are not. As the equal-frequency and complex-frequency Pais-Uhlenbeck models show, one cannot tell whether a Hamiltonian might be Hermitian just by superficial inspection. One needs to solve the theory first and see what the eigenspectrum looks like. Thus one can have Hamiltonians that do not look Hermitian but are similarity equivalent to ones that are Hermitian, and  one can have Hamiltonians that do look Hermitian but are not at all.

As we see from these examples,  whether or not an action is $CPT$ symmetric is an intrinsic property of the unconstrained action itself prior to any stationary variation, but whether or not a Hamiltonian is Hermitian is a property of the stationary solution alone.\footnote{While one can construct the Hamiltonian from the energy-momentum tensor, the energy-momentum tensor is only conserved in solutions to the equations of motion. Hermiticity is thus tied to the solutions to the theory in a way that $CPT$ is not.} Hermiticity of a Hamiltonian cannot be assigned a priori, and can only be determined after the theory has been solved. However, the $CPT$ properties of actions or fields can be assigned a priori (i.e. prior to a functional variation of the action, and thus a property of every variational path and not just the stationary one), and thus that is how Hamiltonians and fields should be characterized. One cannot write down any $CPT$ invariant theory that up to similarity transformations does not have the same form as a Hermitian theory, though whether any such $CPT$ invariant Hamiltonian actually is similarity equivalent to a Hermitian one is only establishable by constructing the solutions to the theory and cannot be determined ahead of time.

Turning now to the study of \cite{Alexandre2018}, we note that it displays all of the features that we have just described. The interest of the authors of \cite{Alexandre2018} was in exploring the status of the Goldstone theorem in non-Hermitian but $PT$-symmetric  theories, and so they took as an example a relativistic field theory whose action was not Hermitian, i.e. not Hermitian by superficial inspection. However, by a similarity transformation it could be brought to a form given in (\ref{GPT46}) in which the action is Hermitian by superficial inspection (i.e.  no factors of $i$ and operators that are presumed to be Hermitian). However, while it now appears to be Hermitian it could not be since in the tree approximation that they studied the ensuing mass matrix was not Hermitian either. With the mass matrix having the three possible $PT$ symmetry realizations (real and unequal eigenvalues, real and equal eigenvalues, eigenvalues in complex conjugate pairs) for various values of its parameters, the tree approximation  to the model of \cite{Alexandre2018} completely parallels the discussion of the three realizations of Pais-Uhlenbeck two-oscillator model given in (\ref{GPT54}), (\ref{GPT55}) and (\ref{GPT56}), where the Hamiltonian looks to be Hermitian but is not. It is of interest to note that to establish that the Pais-Uhlenbeck two-oscillator model theory is not Hermitian we had to construct wave functions and examine there asymptotic behavior, while for the tree approximation to the model of \cite{Alexandre2018} we only need to look at a finite-dimensional matrix. Thus we can start with a fully-fledged field theory such as that based on the action given in (\ref{GPT1}), (\ref{GPT46}) or (\ref{GPT51}) and not need to identify the region in the complex plane where the functional path integral might exist or need to descend to the quantum mechanical limit  and look at the asymptotic behavior of wave functions in order to determine whether or not the theory is Hermitian.\footnote{These issues would only start to come up in fluctuations around the tree approximation minimum, with a one loop calculation having been provided in \cite{Alexandre2018}.} In the broken symmetry case we only need look at the finite-dimensional mass matrix that we get in tree approximation.

For parameters in the model of \cite{Alexandre2018} that obey $(2m_1^2m_2^2-m_2^4-3\mu^4)^2-4\mu^4m_2^4>0$, the mass matrix can be brought to a Hermitian form by the similarity transformation presented in (\ref{GPT22}). Thus in this case the mass matrix is Hermitian in disguise. For this particular example the Goldstone theorem is the standard one, since if one can derive the Goldstone theorem in a Hermitian theory, it continues to hold if one makes a similarity transformation on it.\footnote{Technically, that would have automatically been the case if the authors of \cite{Alexandre2018} had used a conventional variational procedure, though in fact they did not. It is however the case for the study that we have presented here.} Whether or not the mass matrix given in  (\ref{GPT11}) actually can be transformed to a Hermitian matrix  depends on the values of the parameters in the action. However, as we have seen, no matter what the values of these parameters, and no matter whether the $CPT$-invariant mass matrix is realized by real eigenvalues, complex pairs of eigenvalues, or is of  Jordan-block form, for any choice of the parameters one is able to obtain a Goldstone theorem. One can thus anticipate a Englert-Brout-Higgs mechanism for a local extension of the continuous symmetry that we have broken spontaneously, and we turn now to this issue.

\section{Spontaneously Broken non-Hermitian Theory with a Continuous Local Symmetry} 
\label{S6}

Now that we have seem that we can consistently implement the Goldstone mechanism in a $CPT$-symmetric, non-Hermitian theory, it is natural to ask whether we can also implement the familiar Englert-Brout-Higgs mechanism developed in \cite{Englert1964,Higgs1964a,Higgs1964b,Guralnik1964}. To this end we introduce a local gauge invariance and a gauge field $A_{\mu}$, and with $F_{\mu\nu}=\partial_{\mu}A_{\nu}-\partial_{\nu}A_{\mu}$ replace (\ref{GPT1}) and (\ref{GPT3})  by
\begin{eqnarray}
I(\phi_1,\phi_2,\phi^*_1,\phi^*_2,A_{\mu})&=&\int d^4x\bigg{[}(-i\partial_{\mu}+eA_{\mu})\phi^*_1(i\partial^{\mu}+eA^{\mu})\phi_1+(-i\partial_{\mu}+eA_{\mu})\phi^*_2(i\partial^{\mu}+eA^{\mu})\phi_2
\nonumber\\
&+&m_1^2\phi_1^*\phi_1-m_2^2\phi^*_2\phi_2-\mu^2(\phi^*_1\phi_2-\phi^*_2\phi_1)-\frac{g}{4}(\phi^*_1\phi_1)^2 -\frac{1}{4}F_{\mu\nu}F^{\mu\nu}\bigg{]},
\label{GPT62}
\end{eqnarray}
and 
\begin{eqnarray}
\phi_1\rightarrow e^{i\alpha(x)}\phi_1,\quad \phi^*_1\rightarrow e^{-i\alpha(x)}\phi^*_1,\quad \phi_2\rightarrow e^{i\alpha(x)}\phi_2,\quad \phi^*_2\rightarrow e^{-i\alpha(x)}\phi_2,\quad
eA_{\mu}\rightarrow eA_{\mu}+\partial_{\mu}\alpha(x).
\label{GPT63}
\end{eqnarray}
With (\ref{GPT2}), the $I(\phi_1,\phi_2,\phi^*_1,\phi^*_2,A_{\mu})$ action is $CPT$ invariant since both  $i$ and $A_{\mu}$ are $CPT$ odd (spin one fields have odd  $CPT$ \cite{Weinberg1995}).

We make the same decomposition of $\phi_1$ and $\phi_2$ fields as in (\ref{GPT5}), and replace (\ref{GPT6}) by 
\begin{eqnarray} 
I(\chi_1,\chi_2,\psi_1,\psi_2,A_{\mu})&=&\int d^4x \bigg{[}\frac{1}{2}\partial_{\mu}\chi_1\partial^{\mu}\chi_1+
\frac{1}{2}\partial_{\mu}\chi_2\partial^{\mu}\chi_2+
\frac{1}{2}\partial_{\mu}\psi_1\partial^{\mu}\psi_1+
\frac{1}{2}\partial_{\mu}\psi_2\partial^{\mu}\psi_2+
\frac{1}{2}m_1^2(\chi_1^2+\chi_2^2)
\nonumber\\
&-&
\frac{1}{2}m_2^2(\psi_1^2+\psi_2^2)-
i\mu^2(\chi_1\psi_2-\chi_2\psi_1)
-\frac{g}{16}(\chi_1^2+\chi_2^2)^2 
\nonumber\\
&-&eA^{\mu}\left(\chi_1\partial_{\mu}\chi_2-\chi_2\partial_{\mu}\chi_1+
\psi_1\partial_{\mu}\psi_2-\psi_2\partial_{\mu}\psi_1\right)
\nonumber\\
&+&\frac{e^2}{2}A_{\mu}A^{\mu}\left[\chi_1^2+\chi_2^2+\psi_1^2+\psi_2^2\right]-\frac{1}{4}F_{\mu\nu}F^{\mu\nu}
\bigg{]},
\label{GPT64}
\end{eqnarray}
In the tree approximation minimum used above in which  $(g/4)\bar{\chi}_1^2=m_1^2-\mu^4/m_2^2$, $\bar{\psi}_2=-i\mu^2\bar{\chi}_1/m_2^2$, $\hat{\chi}_2=0$, $\hat{\psi}_1=0$, we induce a mass term for $A_{\mu}$ of the form
\begin{eqnarray}
m^2(A_{\mu})=e^2\left(\bar{\chi}_1^2+\bar{\chi}_2^2+\bar{\psi}_1^2+\bar{\psi}_2^2\right)
=e^2\bar{\chi}_1^2\left(1-\frac{\mu^4}{m_2^4}\right)
=\frac{4e^2}{g}\frac{(m_1^2m_2^2-\mu^4)(m_2^4-\mu^4)}{m_2^6}.
\label{GPT65}
\end{eqnarray}

However, before assessing the implications of (\ref{GPT65}) we recall that in Sec. \ref {S4} we had to reconcile $I(\chi_1,\chi_2,\psi_1,\psi_2)$ with the Hermiticity concern raised in \cite{Alexandre2018}. The same is now true of $I(\chi_1,\chi_2,\psi_1,\psi_2,A_{\mu})$. In Sec. \ref{S4} we had identified three solutions for $I(\chi_1,\chi_2,\psi_1,\psi_2)$, and all can be implemented for $I(\chi_1,\chi_2,\psi_1,\psi_2,A_{\mu})$. Thus we can consider a judicious choice of which fields are Hermitian and which are anti-Hermitian,  a judicious choice of which fields are $CPT$ even and which are $CPT$ odd, or can apply similarity transformations that generate complex phases that affect both Hermiticity and $CPT$ parity. 

In regard to Hermiticity, if we take $A_{\mu}$ to be Hermitian (i.e. complex conjugate even), and as before take $\psi_1$ and $\psi_2$ to be anti-Hermitian (complex conjugate odd), then $I(\chi_1,\chi_2,\psi_1,\psi_2,A_{\mu})$ will be invariant under complex conjugation, as will then be the equations of motion and tree approximation minimum that follow from it, and  (\ref{GPT65}) will hold. Also, as we had noted in Sec. \ref{S5}, even though $I(\chi_1,\chi_2,\psi_1,\psi_2,A_{\mu})$ might now be invariant under complex conjugation it does not follow that the scalar field mass matrix $M$ given in (\ref{GPT9}) has to be Hermitian, and indeed it is not. 

If we transform $\psi_1$ and $\psi_2$ as in (\ref{GPT41}) but make no transformation on $A_{\mu}$, we obtain 
\begin{eqnarray}
S(\psi_1)S(\psi_2)I(\chi_1,\chi_2,\psi_1,\psi_2,A_{\mu})S^{-1}(\psi_2)S^{-1}(\psi_1)=I^{\prime}(\chi_1,\chi_2,\psi_1,\psi_2)
\label{GPT66}
\end{eqnarray}
where
\begin{eqnarray}
I^{\prime}(\chi_1,\chi_2,\psi_1,\psi_2,A_{\mu})&=&\int d^4x \bigg{[}\frac{1}{2}\partial_{\mu}\chi_1\partial^{\mu}\chi_1+
\frac{1}{2}\partial_{\mu}\chi_2\partial^{\mu}\chi_2-
\frac{1}{2}\partial_{\mu}\psi_1\partial^{\mu}\psi_1-
\frac{1}{2}\partial_{\mu}\psi_2\partial^{\mu}\psi_2+
\frac{1}{2}m_1^2(\chi_1^2+\chi_2^2)
\nonumber\\
&+&
\frac{1}{2}m_2^2(\psi_1^2+\psi_2^2)-
\mu^2(\chi_1\psi_2-\chi_2\psi_1)
-\frac{g}{16}(\chi_1^2+\chi_2^2)^2 
\nonumber\\
&-&eA^{\mu}\left(\chi_1\partial_{\mu}\chi_2-\chi_2\partial_{\mu}\chi_1-
\psi_1\partial_{\mu}\psi_2+\psi_2\partial_{\mu}\psi_1\right)
\nonumber\\
&+&\frac{e^2}{2}A_{\mu}A^{\mu}\left(\chi_1^2+\chi_2^2-\psi_1^2-\psi_2^2\right)-\frac{1}{4}F_{\mu\nu}F^{\mu\nu}
\bigg{]},
\label{GPT67}
\end{eqnarray}
As constructed, $ I^{\prime}(\chi_1,\chi_2,\psi_1,\psi_2,A_{\mu})$ is invariant under complex  conjugation if $\psi_1$ and $\psi_2$ are even under complex conjugation. Now the tree approximation minimum is given by 
$(g/4)\bar{\chi}_1^2=m_1^2-\mu^4/m_2^2$, $\bar{\psi}_2=\mu^2\bar{\chi}_1/m_2^2$, $\hat{\chi}_2=0$, $\hat{\psi}_1=0$, $A_{\mu}=0$, and we induce a mass term for $A_{\mu}$ of the form
\begin{eqnarray}
m^2(A_{\mu})=e^2\left(\bar{\chi}_1^2+\bar{\chi}_2^2-\bar{\psi}_1^2-\bar{\psi}_2^2\right)
=e^2\bar{\chi}_1^2\left(1-\frac{\mu^4}{m_2^4}\right)
=\frac{4e^2}{g}\frac{(m_1^2m_2^2-\mu^4)(m_2^4-\mu^4)}{m_2^6}.
\label{GPT68}
\end{eqnarray}
The mass of the gauge boson is thus again given by (\ref{GPT65}). Finally, in regard to interpreting the star symbol as the $CPT$ conjugate, since $A_{\mu}$ is real there is no change in the  discussion presented in Sec. \ref{S4},  and (\ref{GPT65}) continues to hold.

As we can see from (\ref{GPT65}) and (\ref{GPT68}), the gauge boson does indeed acquire a non-zero mass unless  $m_1^2m_2^2=\mu^4$ or $m_2^4=\mu^4$. The first of these conditions is not of significance since if $m_1^2m_2^2=\mu^4$ it follows that $\bar{\chi}_1$ (and thus $\bar{\psi}_2$) is zero and there is no symmetry breaking, and the gauge boson stays massless.\footnote{If $m_1^2m_2^2-\mu^4=0$, the unbroken $(\chi_1,\psi_2)$ sector mass matrix given in (\ref{GPT64}) is of the form $(1/2m_2^2)(\mu^2\chi_1-im_2^2\psi_2)^2$. It has two eigenvalues, $\lambda_a=\mu_4/m_2^2-m_2^2$ and $\lambda_b=0$. In the $(\chi_1,\psi_2)$ basis the right-eigenvector for $\lambda_a$ is ($\mu^2,-im_2^2)$, while the right-eigenvector for $\lambda_b$ is ($m_2^2,-i\mu_2^2)$. The fact that $\lambda_b$ is zero  is not of significance since it occurs in the absence of  spontaneous symmetry breaking, and would thus not be maintained under radiative corrections.} However, the condition $m_2^4=\mu^4$ is related to the symmetry breaking since it does not oblige $\bar{\chi}_1$ to vanish. Moreover, since the $m_2^4=\mu^4$  condition does not constrain the $(\tilde{\chi}_1,\tilde{\psi}_2)$ sector $\lambda_{\pm}$ eigenvalues given in (\ref{GPT11}) in any particular way ($m_1^2$ not being constrained by the $m_2^4=\mu^4$ condition), we see that regardless of whether or not $m_2^4$ and $\mu^4$ are in fact equal to each other, we obtain the Englert-Brout-Higgs mechanism in the $(\tilde{\chi}_2,\tilde{\psi}_1)$ sector no matter how the antilinear symmetry is realized in the $(\tilde{\chi}_1,\tilde{\psi}_2)$ sector, be it all eigenvalues real, eigenvalues in a complex pair, or mass matrix being of non-diagonalizable Jordan-block form. In the $(\tilde{\chi}_2,\tilde{\psi}_1)$ sector both $\lambda_0$ and $\lambda_1$ as given in (\ref{GPT11}) are real, and not degenerate with each other as long as $m_2^4\neq \mu^4$. However, something very interesting occurs if $m_2^4=\mu^4$. Then the $(\tilde{\chi}_2,\tilde{\psi}_1)$ sector becomes Jordan block and the Goldstone boson acquires zero-norm. Since the Goldstone boson can no longer be considered to be a normal positive norm particle, it cannot combine with the gauge boson to give the gauge boson a longitudinal component and make it massive. And as we see, and just as required by consistency, in that case the gauge boson stays massless. Thus in a non-Hermitian but $CPT$-symmetric theory it is possible to spontaneously break a continuous local symmetry and yet not obtain a massive gauge boson. 

\section{Summary} 
\label{S7}

In the non-relativistic  antilinear symmetry program one replaces Hermiticity of a Hamiltonian by antilinearity as the guiding principle for quantum mechanics for both infinite-dimensional wave mechanics and either finite- or infinite-dimensional matrix mechanics. For infinite-dimensional relativistic quantum field theories whose actions are invariant under the complex Lorentz group the antilinear symmetry is uniquely prescribed to be $CPT$. Hamiltonians that have an antilinear symmetry can of course be Hermitian as well, with all energy eigenvalues then being real and all energy eigenvectors being complete. However, in general, antilinear symmetry permits two additional options for Hamiltonians that cannot be realized in Hermitian theories, namely energy eigenvalues could be real while energy eigenvectors could be incomplete (Jordan block), or energy eigenvectors could still be complete but energy eigenvalues could come in complex conjugate pairs. In the first case all Hilbert space inner products are positive definite, in the second (Jordan-block) case norms are zero, and in the third case the only norms are transition matrix elements and their values are not constrained to be positive or to even be real. Moreover, in the antilinear symmetry program Hamiltonians that look to be Hermitian by superficial inspection do not have to be, while Hamiltonians that do not look to be Hermitian by superficial inspection can actually be similarity equivalent to Hamiltonians that are Hermitian (viz. Hermitian in disguise). 

In applications of antilinear symmetry to relativistic systems it is of interest to see how many standard results that are obtained in Hermitian theories might still apply in the non-Hermitian case, and whether one could obtain new features that could not be realized in the Hermitian case. To address these issues Alexandre, Ellis, Millington and Seynaeve studied how spontaneously broken symmetry ideas and results translate to the non-Hermitian environment. With broken symmetry and the possible existence of massless Goldstone bosons being intrinsically relativistic concepts, they explored a $CPT$ symmetric but non-Hermitian two complex scalar field relativistic quantum field theory with a continuous global symmetry. Their actual treatment of the problem was somewhat unusual in that they allowed for non-vanishing surface terms to contribute in the functional variation of the action, with this leading to a specific non-standard set of equations of motion of the fields. Their reason for doing this was that the equations of motion obtained by the standard variational procedure with endpoints held fixed were not invariant under complex conjugation. However, they still found a broken symmetry solution with an explicit massless Goldstone boson.  

In the treatment of the same model that we provide here  we use a conventional variational calculation in which fields are held fixed at the endpoints. However, to get round the complex conjugation difficulty we make a similarity transformation on the fields which then allows us to be able to maintain invariance of the equations of motion under complex conjugation (the similarity transformation itself being complex). However, on doing this we obtain an action that appears to be Hermitian, and if it indeed were to be Hermitian there would be nothing new to say about broken symmetry that had not already been said for Hermitian theories. However, while appearing to be Hermitian the theory in fact is not, and thus it does fall into the non-Hermitian but antilinearly symmetric category.

In their analysis Alexandre, Ellis, Millington and Seynaeve studied  the tree approximation to the equations of motion of the theory and found broken symmetry solutions. What is particularly noteworthy of their analysis is that even though they were dealing with a fully-fledged infinite-dimensional quantum field theory, in the tree approximation the mass matrix that was needed to determine whether there might be any massless Goldstone boson was only four dimensional (viz. the same number as the number of independent fields in the two complex scalar field model that they studied). As such, the mass matrix that they obtained is not Hermitian, and given  the underlying antilinear $CPT$ symmetry of the model and thus of the mass matrix,  the mass matrix  is immediately amenable to the full apparatus of the antilinear symmetry program as that apparatus holds equally for fields and matrices. Alexandre, Ellis, Millington and Seynaeve studied just one realization of the antilinear symmetry program, namely the one where all eigenvalues of the mass matrix are real and the set of all of its eigenvectors is complete. In our analysis we obtain the same mass matrix (which we must of course since all we have done is make a similarity transformation on their model), and show that in this particular realization the mass matrix can be brought to a Hermitian form by a similarity transformation, to thus be Hermitian in disguise.

However, this same mass matrix admits of the two other realizations of antilinear symmetry as well, namely the non-diagonalizable Jordan-block case and the complex conjugate eigenvalue pair case. And in all of these cases we show that there is a massless Goldstone boson. In this regard the Jordan-block case is very interesting because it permits the Goldstone boson itself to be one of the zero norm states that are characteristic of Jordan-block matrices. That these cases can occur at all is because while the similarity transformed action that we use appears to be Hermitian it actually is not, something however that one cannot ascertain without first solving the theory. Finally, we extend the model to a local continuous symmetry by introducing a massless gauge boson, and find that the massless Goldstone boson can be incorporated into the massless gauge boson and make it massive by the Englert-Brout-Higgs mechanism in all realizations of the antilinear symmetry except one, namely the Jordan-block Goldstone mode case. In that case we find that since the Goldstone boson then has zero norm, it does not get incorporated into the gauge boson, with the gauge boson staying massless. In this case we have a spontaneously broken local gauge symmetry and yet do not get a massive gauge boson. This option cannot be obtained in the standard Hermitian case where all states have positive norm, to thus show how rich the non-Hermitian antilinear symmetry program can be.

\appendix

\setcounter{equation}{0}
\def\theequation{A\arabic{equation}}

\section{Meaning of the Tree Approximation in the non-Hermitian Case}
\label{App1}

In general the fields used in the tree approximation to a quantum field theory are c-number matrix elements of q-number quantum fields. Given the left- and right-eigenstates introduced in Sec. \ref{S3} we can identify which particular states are involved in (\ref{GPT8}). Specifically, we can identify the c-number tree approximation fields as the matrix elements $\langle \Omega_L|\phi|\Omega_R\rangle=\langle \Omega_R|V\phi|\Omega_R\rangle$, i.e. c-number matrix elements of the q-number fields between the left and right vacua. 

In quantum field theory one introduces a generating functional via the Gell-Mann-Low adiabatic switching method. The discussion in the Hermitian case is standard and of ancient vintage, and following the convenient discussion in \cite{Mannheim2017}, here we adapt it to the non-Hermitian case. In the adiabatic switching procedure one introduces a quantum-mechanical Lagrangian density $L_0$ of interest, switches on a real local c-number source $J(x)$ for some quantum field $\phi(x)$ at time $t=-\infty$, and switches $J(x)$  off at $t=+\infty$. While the source is active the Lagrangian density of the theory is given by $L_J=L_0+J(x)\phi(x)$. Before the source is switched on the system is in the ground state of the Hamiltonian $H_0$ associated with $L_0$ with right-eigenvector $|\Omega_R^-\rangle$ and  left-eigenvector $\langle \Omega_L^-|=\langle \Omega_R^-|V$. (Here $V$ implements $VH_0V^{-1}=H_0^{\dagger}$ and is independent of $J$.) And after the source is switched off the system is in the state with right-eigenvector $|\Omega_R^+\rangle$ and left-eigenvector $\langle \Omega_L^+|=\langle \Omega_R^+|V$. ($V$ again implements $VH_0V^{-1}=H_0^{\dagger}$.) While $|\Omega_R^-\rangle$ and $|\Omega_R^+\rangle$ are both eigenstates of  $H_0$, they differ by a phase, a phase that is fixed by $J(x)$ according to 
\begin{eqnarray}
\langle \Omega_L^+|\Omega_R^-\rangle |_J=\langle \Omega_R^+|V|\Omega_R^-\rangle |_J=\langle \Omega_L^J|T\exp\left[i\int d^4x(L_0+J(x)\phi(x))\right]|\Omega_R^J\rangle=e^{iW(J)},
\label{A1}
\end{eqnarray}
as written in terms of the vacua when $J$ is active. This expression serving to define the functional $W(J)$, with $W(J)$ serving as the generator of the connected $J=0$ theory Green's functions 
\begin{eqnarray}
G^{n}_0(x_1,...,x_n)=\langle\Omega_L|T[\phi(x_1)...\phi(x_n)]|\Omega_R\rangle
\label{A2}
\end{eqnarray}
according to 
\begin{eqnarray}
W(J)=\sum_n\frac{1}{n!}\int d^4x_1...d^4x_nG^{n}_0(x_1,...,x_n)J(x_1)...J(x_n).
\label{A3}
\end{eqnarray} 
Given $W(J)$, via functional variation we can construct the so-called classical (c-number) field $\phi_C(x)$ 
\begin{eqnarray}
\phi_C(x)=\frac{\delta W}{\delta J(x)}=\frac{\langle \Omega_L^+|\phi(x)|\Omega_R^-\rangle}{\langle \Omega_L^+|\Omega_R^-\rangle}\bigg{|}_J
\label{A4}
\end{eqnarray}
and the effective action functional
\begin{eqnarray}
\Gamma(\phi_C)=W(J)-\int d^4x J(x)\phi_C(x)=\sum_n\frac{1}{n!}\int d^4x_1...d^4x_n\Gamma^{n}_0(x_1,...,x_n)\phi_C(x_1)...\phi_C(x_n),
\label{A5}
\end{eqnarray}
with the $\Gamma^{n}_0(x_1,...,x_n)$ being the one-particle-irreducible, $\phi_C=0$, Green's functions of the quantum field $\phi(x)$.  Functional variation of $\Gamma(\phi_C)$ then yields
\begin{eqnarray}
\frac{\delta \Gamma(\phi_C)}{\delta \phi_C}=\frac{\delta W}{\delta J }\frac{\delta J}{\delta \phi_C}-J-\frac{\delta J}{\delta \phi_C} \phi_C=-J,
\label{A6}
\end{eqnarray}
to relate $\delta \Gamma(\phi_C)/\delta \phi_C$ back to the source $J$.

On expanding in momentum space around the point where all external momenta vanish, we can write $\Gamma(\phi_C)$ as
\begin{eqnarray}
\Gamma(\phi_C)=\int d^4x\left[-V(\phi_C)+\frac{1}{2}Z(\phi_C)\partial_{\mu}\phi_C\partial^{\mu}\phi_C+....\right].
\label{A7}
\end{eqnarray}
The quantity 
\begin{eqnarray}
V(\phi_C)=\sum_n\frac{1}{n!}\Gamma^{n}_0(q_i=0)\phi_C^n
\label{A8}
\end{eqnarray}
is known as the effective potential as introduced in \cite{Goldstone1962,Jona-Lasinio1964} (a potential that is spacetime independent if $\phi_C$ is), while the $Z(\phi_C)$ term  serves as the  kinetic energy of $\phi_C$. 

The significance of $V(\phi_C)$ is that when $J$ is zero and $\phi_C$ is spacetime independent, we can write $V(\phi_C)$ as 

\begin{eqnarray}
V(\phi_C)=\frac{1}{V}\left(\langle S_L|H_0|S_R\rangle-\langle N_L|H_0|N_R\rangle\right)
\label{A9}
\end{eqnarray}
in a volume $V$, where $|S_R\rangle$ and $|N_R\rangle$ are spontaneously broken and normal vacua in which $\langle S_L|\phi|S_R\rangle$ is nonzero and $\langle N_L|\phi|N_R\rangle$ is zero. The search for non-trivial tree approximation minima is then a search for states $|S_R\rangle$ in which $V(\phi_C)$ would be negative. In the non-Hermitian case then the $V(\phi_C)$ associated with left and right vacua is the needed effective potential.\footnote{In addition we note that once one we have identified $\langle\Omega_L|T[\phi(x_1)...\phi(x_n)]|\Omega_R\rangle$ as the connected Green's functions that are relevant  in the non-Hermitian case, we can write them as path integrals, and we refer the reader to \cite{Mannheim2013,Mannheim2018,Alexandre2018} for further details.}

In reference to the Goldstone theorem, we note that in writing down Ward identities one begins with operator relations for time-ordered products of general fields and current operators of the generic form 
\begin{eqnarray}
\partial_{\mu}\left[T\left(j^{\mu}(x)A(0)\right)\right]=\delta(x^0)[j^{0}(x),A(0)]+T
\left(\partial_{\mu} j^{\mu}A(0)\right),
\label{A10}
\end{eqnarray}
where $A(0)$ is a product  of fields at the origin of coordinates.  We restrict to  the case where $\partial_{\mu}j^{\mu}(x)=0$, and  take matrix elements in the vacuum (normal or spontaneously broken), only unlike in the Hermitian case in the non-Hermitian case we take matrix elements in the left- and right-vacua. Since there is only one four-momentum $p^{\mu}$ in the problem in Fourier space we can set
\begin{eqnarray}
\langle \Omega_L|T(j^{\mu}(x)B(0))|\Omega_R\rangle=\frac{1}{(2\pi)^4}\int d^4p e^{ip\cdot x}p^{\mu}F(p^2),
\label{A11}
\end{eqnarray}
where $F(p^2)$ is a scalar function. On introducing $Q(t)=\int d^3x j^0(x)$, we integrate both sides of (\ref{A11}) with $\int d^4x$. This gives
\begin{eqnarray}
i\int d^4p\delta^4(p)p^2F(p)=\langle \Omega_L|[Q(t=0),B(0)]|\Omega_R\rangle.
\label{A12}
\end{eqnarray}
Should the right-hand side of (\ref{A12}) not vanish (i.e. $Q(t=0)|\Omega_R\rangle \neq 0$ or $\langle \Omega_L|Q(t=0) \neq 0$), there would then have to be  a massless pole at $p^2=0$ on the left-hand side. This then is the Goldstone theorem, as adapted to the non-Hermitian case. As we see, by formulating non-Hermitian theories in terms of left- and right-eigenvectors, the extension of the discussion of spontaneously broken symmetry to the non-Hermitian case is straightforward.

The specific structure of Ward identities such as that given in (\ref{A12}) only depends on the symmetry behavior associated with the currents of interest. Since relations such as (\ref{A10}) are operator identities they hold independent of the states in which one calculates matrix elements of them.  In the non-Hermitian but $CPT$-symmetric situation,  in order to look for a spontaneous breaking of the continuous global symmetry associated with the currents of interest  one takes matrix elements of the relevant Ward identity in the $\langle S_L|$ and $|S_R\rangle$ states,  and as discussed in \cite{Alexandre2018}, one looks to see if the consistency of the Ward identity matrix elements in those states requires the existence of massless Goldstone bosons. In regard to the spontaneous breakdown of a continuous local symmetry in the non-Hermitian but $CPT$-symmetric case, the authors of \cite{Alexandre2018} had left open the question of whether one could achieve the Englert-Brout-Higgs mechanism if one uses their non-standard variational procedure.\footnote{After we had finished this paper Alexandre, Ellis, Millington and Seynaeve released a follow up paper ``Gauge invariance and the Englert-Brout-Higgs mechanism in non-Hermitian field theories", arXiv:1808.00944 [hep-th] \cite{Alexandre2018a}. In it the authors extend the analysis of \cite{Alexandre2018} to the Englert-Brout-Higgs mechanism. Their paper can be regarded as complementary to our own.} Since we use a standard variational procedure and standard Noether theorem approach and continue to use the same $\langle S_L|$ and $|S_R\rangle$ states, we can readily extend our approach to the local symmetry case. And  we find that in all realizations of the antilinear symmetry we can achieve the Englert-Brout-Higgs mechanism just as in the standard Hermitian case, save only for the particular Jordan-block situation in which the Goldstone boson itself has zero norm, a case in which, despite the spontaneous symmetry breaking, the gauge boson stays massless.

\end{document}